\documentclass[12pt,preprint]{aastex}

\newcommand{\gcc}{\ \mathrm{g\ cm^{-3} }}
\newcommand{\micros}{\ \mu s}
\newcommand{\cms}{\ \mathrm{cm \ s^{-1}}}

\usepackage{epsf,color,graphicx}

\epsfverbosetrue

\begin{document} 

\title{Mapping Initial Hydrostatic Models in Godunov Codes}

\shorttitle{Hydrostatic Models in Godunov Codes}

\author{M.~Zingale\altaffilmark{1}, 
	L.~J.~Dursi\altaffilmark{2,3},
        J.~ZuHone\altaffilmark{4},
        A.~C.~Calder\altaffilmark{2,3},
        B.~Fryxell\altaffilmark{2,3},
        T.~Plewa\altaffilmark{2,3,5}, 
        J.~W.~Truran\altaffilmark{2,3},
        A.~Caceres\altaffilmark{2,3},
        K.~Olson\altaffilmark{2,3,6},
        P.~M.~Ricker\altaffilmark{2,3},
	K.~Riley\altaffilmark{2,3},
        R.~Rosner\altaffilmark{2,3},
        A.~Siegel\altaffilmark{2,3},
        F.~X.~Timmes\altaffilmark{2,3}, 
        N.~Vladimirova\altaffilmark{2,3}}

\altaffiltext{1}{Dept.\ of Astronomy \& Astrophysics,
                 The University of California, Santa Cruz,
                 Santa Cruz, CA 95064}

\altaffiltext{2}{Dept.\ of Astronomy \& Astrophysics, 
                 The University of Chicago, 
                 Chicago, IL  60637}

\altaffiltext{3}{Center for Astrophysical Thermonuclear Flashes, 
                 The University of Chicago, 
                 Chicago, IL  60637}

\altaffiltext{4}{Dept.\ of Physics,
                 University of Illinois, Urbana-Champaign,
                 Urbana, IL 61801}

\altaffiltext{5}{Nicolaus Copernicus Astronomical Center,
                 Bartycka 18,
                 00716 Warsaw, Poland}

\altaffiltext{6}{UMBC/GEST Center, NASA/GSFC, 
                 Greenbelt, MD 20771}

\begin{abstract}

We look in detail at the process of mapping an astrophysical initial
model from a stellar evolution code onto the computational grid of an
explicit, Godunov type code while maintaining hydrostatic equilibrium.
This mapping process is common in astrophysical simulations, when it
is necessary to follow short-timescale dynamics after a period of long
timescale buildup.  We look at the effects of spatial resolution,
boundary conditions, the treatment of the gravitational source terms
in the hydrodynamics solver, and the initialization process itself.
We conclude with a summary detailing the mapping process that yields
the lowest ambient velocities in the mapped model.

\end{abstract}


\newcommand{\avg}[1]{\left<#1\right>}
\newcommand{\avgsub}[2]{\left<#1\right>_{#2}}
\section{Introduction}
Many astrophysical phenomena involve a dramatic change between
timescales of interest --- the slow convection and simmering in the
interior of a white dwarf followed by ignition of a Type Ia supernova,
for example,
or the accretion of a layer of fuel onto a white dwarf or neutron star
leading to ignition and runaway at the base of the layer, producing a
nova or an X-ray burst.  These two regimes are difficult to follow
with a single hydrodynamic algorithm because of the disparity of the
relevant timescales.  Modeling these long timescale events requires
an implicit or anelastic hydrodynamic method; short timescale events
require explicit hydrodynamic methods that can capture the transient
phenomena.  Often, a one-dimensional stellar hydrodynamics code is
used to follow the accretion process until just before the fuel
reaches the ignition temperature.  Simulations of the flash following
the ignition requires a multidimensional hydrodynamics code.  Matching
the two regimes is a difficult process, and can introduce numerous
errors into a calculation.

Simulations in the atmosphere or interior of a star or compact object
often begin with an initial model that was generated by a 1-d implicit
code (see for example \citealt{kepler,glasner,sugimoto,tycho,nova}), that
evolves the system through the long timescale processes (accretion,
slow convection, simmering of nuclear fuel) until just before the
short timescale dynamics begin.  This 1-d initial model is then used
as input to a multidimensional, explicit hydrodynamics code (see for
example \citealt{glasner, kercek, zingale, bazan, kane}).  The mapping
of a 1-d hydrostatic initial model onto a multidimensional grid is the
focus of the present paper.  In the absence of any perturbations or
external forces, the system should remain in hydrostatic equilibrium
after the mapping.

The mapping process can introduce a variety of errors.  It is common
for the two codes to use different equations of state (EOS), which can
have a large effect on the structure of the atmospheres.  Even if the
basic physical components are the same, the treatment of physical
details (for example, Coulomb corrections or ionization) may differ,
leading to differences in the pressure of a fluid element between the
two codes, even with the same density, temperature, and composition.
Simply updating the thermodynamics of the initial model with the new
EOS will most likely push it out of hydrostatic equilibrium.
One-dimensional initial models are almost always created using mixing
length theory to describe the convection, implicitly assuming a
velocity field necessary to transport the energy as required.  If the
1-d model was convectively unstable, it is unclear how to define the
2- or 3-d velocity field consistent with the 1-d input velocity field.
When mapping into higher dimensions, there is not enough information
to set the velocities properly.  Typically the velocities are zeroed
during this transition.

Finally, it is unusual for the number of points and the grid spacing in the
initial model to match that in the multidimensional
grid.  The initial model may have come from a Lagrangian code and
will need to be converted into an Eulerian coordinate system via
\begin{equation}
dm = 4 \pi r^2 \rho(r) dr.
\end{equation}
The initial model will then need to be interpolated onto the new grid,
which will introduce even more hydrostatic equilibrium errors.

Once the model is on the new grid, differences in the hydrodynamical
algorithms can cause problems.  Although the two codes are solving the
same equations, it is also very likely that the discretization used in
the two codes is different.  The very definition of the quantities on
the grid may also differ; the mapping may proceed from a 1-d implicit
finite-difference code, where the values on the grid may represent
nodal points, to a multidimensional finite-volume code, where the
values represent cell averages.  If everything else were constant
between the codes, the differences in the discretization and the
definition of the variables ({\em{e.g.}} pointwise values vs.\ cell
averages) is enough to upset the hydrostatic equilibrium.  Poor
boundary conditions can drive velocities on the grid, pushing our
initial model out of hydrostatic equilibrium.  Ideally, the boundaries
should match the physics of the initial model and present a smooth
state to the hydrodynamics solver.

In this article, we look at bringing an initial model to hydrostatic
equilibrium by studying the effects of the initialization method, the
boundary conditions, and the solver itself.  An initial model
atmosphere is mapped onto the computational grid, transverse to the
direction of gravity.  Since we are not concerned with perturbing the
model after the mapping, all the calculations presented here will be
one-dimensional for computational efficiency.  We consider three
different initial models---a simple analytic model, a simple
hydrostatic equilibrium model with a complicated EOS, and an initial
model from an implicit stellar hydrodynamics code.  These three
different models will allow us to isolate the importance of the
different parts of the code (initialization, boundary conditions, and
solver itself).  Our goal is to find an optimal configuration that
allows us to hold an initial model from a different hydrodynamics code
in equilibrium in the present code until a perturbation or external
force that we impose disturbs it.  Furthermore, we want the resulting
mapped hydrostatic model to be as close to the original initial model
as possible.  Regions where disturbances have not yet propagated
should remain quiescent for tens or hundreds of dynamical timescales.
The absolute magnitude of the velocity that a simulation can tolerate
will be problem dependent; however, it must be considerably less than
the velocity of any dynamics we wish to study (e.g., burning
front speed, convective speed, and certainly the sound speed.)

For the hydrodynamics, we chose to use PPM \citep{ppm}, a widely used
Godunov method \citep{godunov}.  PPM solves the Euler equations in
conservative form, using a finite-volume discretization, guaranteeing
conservation.  PPM is a shock-capturing scheme, which is desirable for
the modeling of the rapid transients in the short-timescale regimes we
wish to study.  The implementation of PPM we use is contained in the
FLASH code
\citep{flash}, and is based on PROMETHEUS \citep{prometheus}.  While we
use PPM to demonstrate results in this paper, only the discussion in \S
\ref{sec:ppmhse}, where we discuss extensions to the method to better
treat the gravitational source term,  is particular to PPM.
Discussion of initialization and boundary conditions should be relevant
to any Godunov-type method, and perhaps other finite-volume methods.
Although FLASH can use an adaptive mesh, in this study we run all the
simulations on a uniform grid. 

This paper is organized as follows: \S 2 discusses the hydrodynamics
algorithm employed in this study and improvements made to better follow
a hydrostatic atmosphere.  In \S 3 we look at the initial models that
will form the basis of our tests.  \S 4 discusses the different
boundary conditions considered.  In \S 5, we show the results of a
grid of calculations of each of the different initial models, varying
spatial resolution, boundary conditions, and the details of the
hydrodynamics.  Finally, we conclude in \S6.

\section{Hydrodynamics}
Understanding how a simulation code treats the hydrodynamics is
critical to being able to accurately initialize the grid and maintain
an atmosphere in hydrostatic equilibrium.  The Godunov-type methods we
consider here solve the compressible Euler equations of continuity,
\begin{equation}
\label{mass_eq}
   \frac{\partial \rho}{\partial t} + {\bf \nabla} \cdot \rho {\bf v} = 0
\enskip ,
\end{equation}
momentum,
\begin{equation}
\label{momentum_eq}
   \frac{\partial \rho {\bf v}}{\partial t} 
   + {\bf \nabla} \cdot \rho {\bf v v} + {\bf \nabla} {P} 
    = \rho {\bf g}
\enskip ,
\end{equation}
energy,
\begin{equation}
\label{totenergy}
    \frac{\partial \rho E}{\partial t} 
    + {\bf \nabla} \cdot \left (\rho E + P \right ) {\bf v}
    =  \rho {\bf v \cdot g}
\enskip , 
\end{equation}
and species advection,
\begin{equation}
\nonumber
\label{abundance_eq}
    \frac{\partial \rho X_i}{\partial t} + {\bf \nabla} \cdot \rho X_i {\bf v} = 0
\enskip ,
\end{equation}
where $\rho$ is the total mass density, ${\bf v}$ is the velocity, $P$
is the pressure, $E$ is the total specific energy, and $X_i$ is the
mass fraction of species $i$.  In all discussion below, unless
otherwise noted, it is assumed that the gravitational acceleration,
$g$, is spatially constant and negative, and in the ${\bf \hat{z}}$
direction so that ${\bf g} = g {\bf \hat{z}}$.

\subsection{\label{sec:cellavg}Cell-averaged quantities}

The Euler equations are expressed above in their conservation-law form,
\begin{equation}
\label{fv}
\frac{\partial {\bf U} }{\partial t} + {\bf \nabla} \cdot {\bf F}({\bf U}) = \bf{S} (\bf{U}) \enskip ,
\end{equation}
where ${\bf U} = \left ( \rho, \rho v, \rho E \right )^T $ are the
conserved quantities.  Equation (\ref{fv}) is the differential form of
the conservation law.  Godunov methods, and other finite-volume methods, 
solve Eq. (\ref{fv}) in integral form:
\begin{equation}
\frac{1}{V} \int \left \{ \frac{\partial {\bf U}}{\partial t} + 
{\bf \nabla} \cdot {\bf F} ({\bf U}) \right \} dV = \frac{1}{V} \int {\bf S} ({\bf U}) dV \enskip ,
\end{equation}
where $V$ is the cell volume.  Using $\avg{{\bf U}}$ to denote the
cell average of quantity ${\bf U}$, we can rewrite this as
\begin{equation}
\label{eq:fv_integral}
\frac{\partial \avg{{\bf U}}}{\partial t} + \frac{1}{V} \oint {\bf F}({\bf U}) {\bf n} \cdot {\bf dS} = \avg{{\bf S} ({\bf U})} \enskip .
\end{equation}
The integral form has the advantage of being able to deal with
discontinuities (e.g. shocks) because it requires less smoothness of
the data \citep{toro}.  The second term in the expression above is
simply the sum of the fluxes through the boundaries of the cell.
Since whatever flux leaves one cell enters the neighboring cell, any
finite-volume discretization of conservative equations will itself be
conservative to roundoff error.

\begin{figure}

\plotone{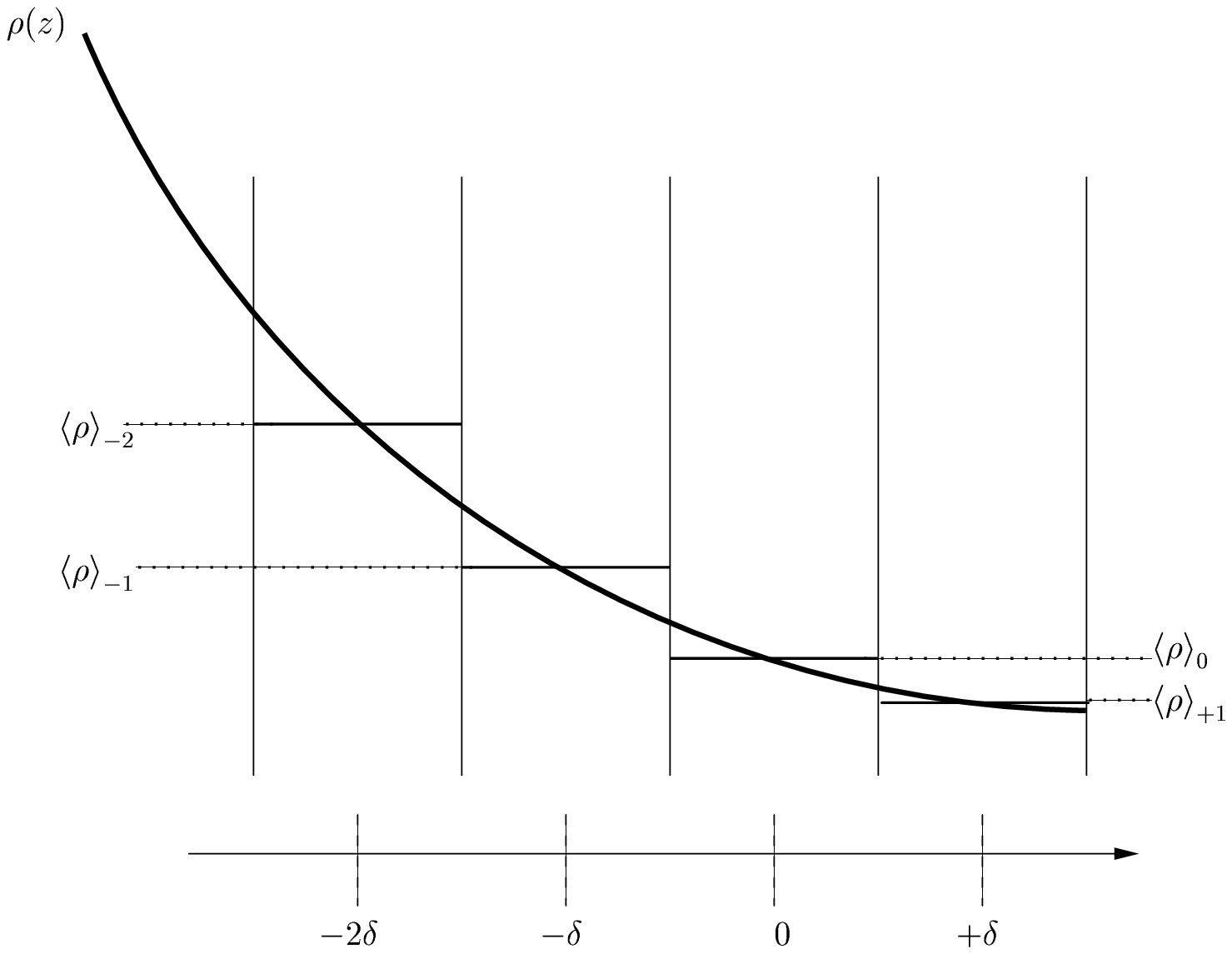}

\caption{\label{interp} A function $\rho_0(z)$ is computed, which
	 reconstructs the data for $\rho$ within zone 0.
	 Shown here is a cubic, as in Eq. (\ref{eq:cubicinterp}); it
	 averages over $z_i = (-2 \delta,-\delta,0,\delta)$ to the
	 previous data values
	 $(\avgsub{\rho}{-2},\avgsub{\rho}{-1},\avgsub{\rho}{0},\avgsub{\rho}{1})$.
	 Reconstructions using lower orders of polynomials need 
         fewer data points.
} \end{figure}

Much of what we will need to do throughout the rest of the paper to bring
a model into hydrostatic equilibrium will involve interpolating or
extrapolating from data to the mesh.  However, because finite-volume
methods deal with cell-averaged quantities, rather than cell-center or
nodal quantities, one cannot use usual interpolants when creating
functions that reconstruct the discrete data.  Rather than using a
function $f(z)$ such that $f\left(z_i\right) = f_i$, one must create a
function such that $({1}/{\delta z}) \int_{z^-}^{z^+} f(z') dz = \left
< f\right>_i$ (see Figure \ref{interp}).  For a uniform mesh in one
dimension, polynomial reconstruction functions with $z=0$ taken to be
the cell center containing the value $\left < f \right>_0$ (see also
\citealt{laney}) are for first-order:
\begin{equation}
\label{eq:linearinterp}
f(z) = \frac{\avgsub{f}{+1} - \avgsub{f}{0}}{\delta}z + \avgsub{f}{0} \enskip ,
\end{equation}
for second order:
\begin{equation}
\label{eq:quadinterp}
f(z) = \frac{\avgsub{f}{+1} - 2\avgsub{f}{0} + \avgsub{f}{-1}}{2 \delta^2} z^2 +
\frac{\avgsub{f}{+1} - \avgsub{f}{-1}}{2 \delta} z + 
\frac{-\avgsub{f}{+1} + 26 \avgsub{f}{0} - \avgsub{f}{-1}}{24} \enskip ,
\end{equation}
for third order:
\begin{eqnarray}
\label{eq:cubicinterp}
f(z) & = & A z^3 + B z^2 + C z + D \enskip ,\\
\nonumber A & = & \frac{\avgsub{f}{+1}-3\avgsub{f}{0}+3\avgsub{f}{-1}-\avgsub{f}{-2}}{6\delta^3} \enskip ,\\ 
\nonumber B & = & \frac{\avgsub{f}{+1}-2\avgsub{f}{0}+\avgsub{f}{-1}}{2\delta^2} \enskip ,\\
\nonumber C & = & \frac{7\avgsub{f}{+1}+15\avgsub{f}{0}-27\avgsub{f}{-1}+5\avgsub{f}{-2}}{24\delta} \enskip ,\\
\nonumber D & = & \frac{-\avgsub{f}{+1}+26\avgsub{f}{0}-\avgsub{f}{-1}}{24} \enskip ,
\end{eqnarray}
and, finally, for a fourth order polynomial:
\begin{eqnarray}
\label{eq:quarticinterp}
f(z) & = & A z^4 + B z^3 + C z^2 + D z + E \enskip ,\\
\nonumber A &=&\frac{\avgsub{f}{+1} - 4\avgsub{f}{0} + 6\avgsub{f}{-1} 
               - 4\avgsub{f}{-2} + \avgsub{f}{-3}}{24 \delta^4} \enskip ,\\
\nonumber B &=& \frac{3\avgsub{f}{+1} - 10\avgsub{f}{0} + 12\avgsub{f}{-1} 
               - 6\avgsub{f}{-2} + \avgsub{f}{-3}}{12 \delta^3} \enskip ,\\
\nonumber C &=& \frac{7\avgsub{f}{+1} - 12\avgsub{f}{0} + 2\avgsub{f}{-1} 
               + 4\avgsub{f}{-2} - \avgsub{f}{-3}}{16 \delta^2} \enskip ,\\
\nonumber D &=& \frac{9\avgsub{f}{+1} + 50\avgsub{f}{0} - 84\avgsub{f}{-1} 
               + 30\avgsub{f}{-2} - 5\avgsub{f}{-3}}{48 \delta} \enskip ,\\
\nonumber E &=& \frac{-71\avgsub{f}{+1} + 2044\avgsub{f}{0} - 26\avgsub{f}{-1} 
               - 36\avgsub{f}{-2} + 9\avgsub{f}{-3}}{1920} \enskip .
\end{eqnarray}
These reconstruction functions will be used throughout the paper.

A high-order Godunov method solves the Euler equations in conservation-law
form using Eq.\ (\ref{eq:fv_integral}).   In one approach, the first
step in the calculation of fluxes between the finite-volume zones is to
perform a reconstruction to represent values of the variables
continuously through the zone.  In PPM, this reconstruction step is
particularly involved.  Care is taken to introduce no new maxima/minima
in the polynomial representation of the solution.  Furthermore, near
shocks and discontinuities, the profile is flattened, so as to avoid
oscillations.  The reconstructed profiles are then averaged over the
region that will be `seen' by waves during the next timestep to generate
average values on the left and right side of the interface, to provide
left and right states as input to the Riemann problem.  The Riemann
solver then constructs the fluxes through the boundary, which are then
used to update the cell-averaged solution vector, $\avg{{\bf U}}$.
The Riemann solver we used is based on \citet{colella_glaz}, and can
handle an arbitrary EOS.

\subsection{\label{sec:ppmhse}Hydrostatic equilibrium}

We note that in the absence of any initial velocities,
Eqs. (\ref{mass_eq}) to (\ref{abundance_eq}) reduce to
\begin{equation}
\label{eq:hse_full}
\frac{\partial \rho}{\partial t} = 0, \quad {\bf \nabla} P = \rho g, \quad
\frac{\partial \rho E}{\partial t} = 0, \quad 
\frac{\partial \rho X_i}{\partial t} = 0 \enskip .
\end{equation}
These equations represent the condition of hydrostatic equilibrium
(HSE).  Simulations beginning from initial models satisfying ${\bf
\nabla} P = \rho g$, with zero velocity everywhere should
maintain that profile with time.  If this expression is not satisfied
perfectly, as differenced in the hydrodynamics algorithm, an
acceleration will result.  The non-linear character of these equations
means the resulting velocities are likely to be amplified and disturb
an initially hydrostatic model.

In the operator-split formulation that FLASH and many other codes use,
the hydrodynamics and the gravity operators are not very closely
coupled.  In this case, maintaining HSE relies on the exact
cancellation of two possibly large terms, ${\bf \nabla}{P}$ and $\rho
g$, calculated in two different ways.  Since this cancellation won't
in general be exact, spurious velocities result.  In Godunov-type
solvers, this cancellation must happen in two places: in the highly
nonlinear Riemann solve across each cell interface during the flux
calculation, and in the subsequent update in each zone.  It is the
Riemann solve which is the most problematic, and which we address
here.

\begin{figure}

\plotone{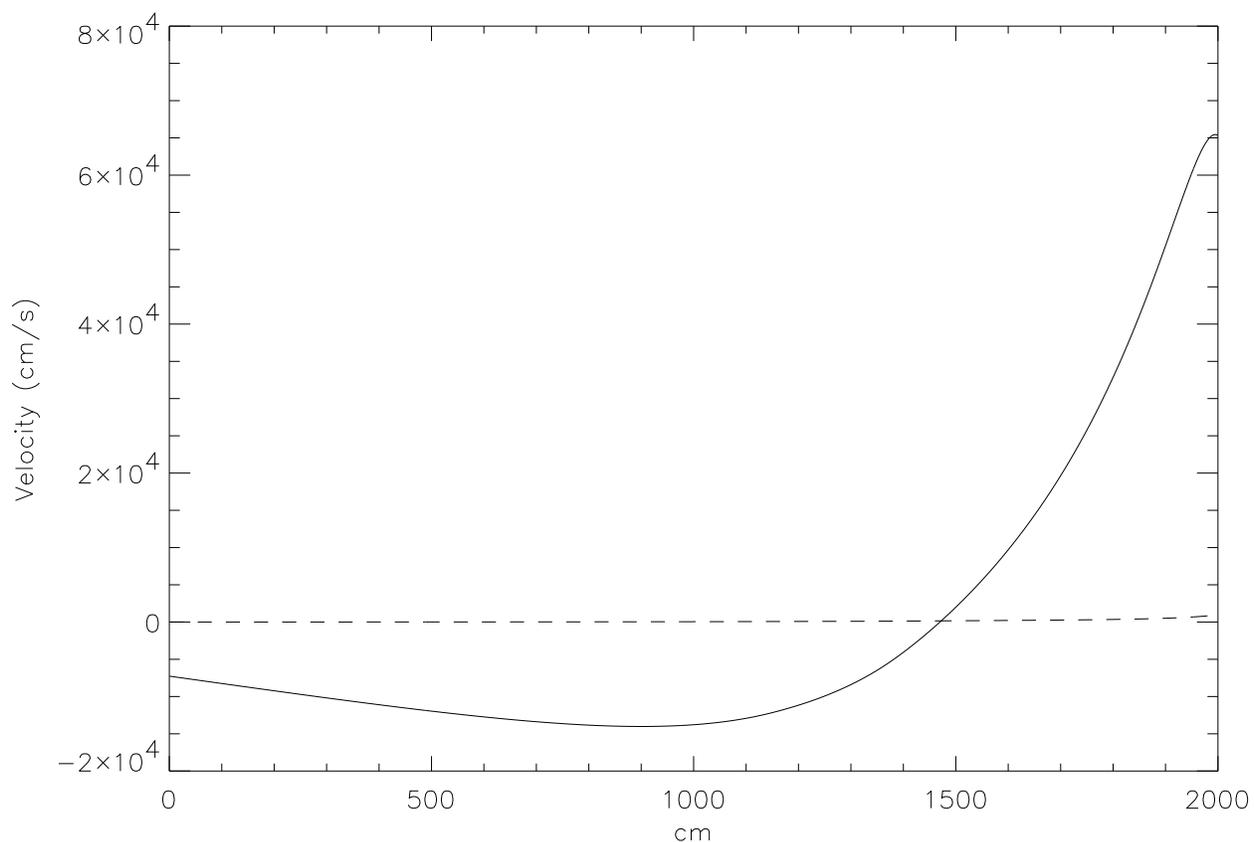}

\caption{\label{fig:lvq_vs_godu} Velocity profile, as a function of 
         vertical position, of the isothermal atmosphere at time
         $t = 2{\mathrm{\mu s}}$ (approximately $1/2$ of a dynamical time)
         when evolved with 3072 points with
         Godunov's method (solid) and Godunov's method corrected
         by LeVeque's method \citep{leveque} (dashed).   The maximum
         velocity in the corrected case, which cannot be seen on the
         scale of the graph, is $975~{\mathrm{cm\,s^{-1}}}$.  The
         atmosphere's pressure scale height ranges from $1036 {\ \mathrm{cm}}$
         at the base to $48 {\ \mathrm{cm}}$ at the top boundary; the
         sound speed ranges from $4.4 \times 10^8 {\ \mathrm{cm\,s^{-1}}}$
         to $9.5 \times 10^7 {\ \mathrm{cm\,s^{-1}}}$.}
\end{figure}

For simulations where one of the acceleration terms dominates, the
resulting small errors may not be significant.  For nearly-hydrostatic
problems, however, the spurious velocities could be disastrous.  Thus,
for these problems, one would like to forgo the splitting of the
imposition of the gravitational and hydrodynamic acceleration, and
instead include the gravitational effects directly in the hydrodynamic
solve.  Other authors, such as \citet{leveque-saasfe}, have suggested
ways of informing Godunov-type hydro solvers of the gravity source
term by notionally putting a constant jump into the energy and density
field given to the Riemann solver to cancel out the source term while
maintaining the cell-average quantities.  This works quite well for
both the Godunov method, where we have implemented the method in the
FLASH code, (see Figure \ref{fig:lvq_vs_godu}) and for a large class
of higher-order Godunov methods \citep{leveque}.

PPM, and other `Reconstruct-Solve-Average' methods
\citep{leveque-saasfe}, use a large stencil to compute a smooth
reconstruction to the fields on either side of the interface, and
average over that reconstruction to generate the left and right states
to the Riemann problem.   Since the range over which the reconstruction
averaged over is determined by the characteristics and isn't known
{\emph{a priori}}, it is not obvious how to nicely and efficiently
compute a constant jump in the states to account for the gravitational
source term.  Nonetheless, a smooth analog is possible.

In our `modified states' version of PPM, when calculating the left and
right states for input to the Riemann solver, we locally subtract off
from the pressure field the pressure that is locally supporting the
atmosphere against gravity; this pressure is unavailable for generating
waves.  At each cell interface ($z = z_{i+1/2}$), we subtract the
pressure locally supporting the atmosphere by computing a
reconstruction to the quantity $(\rho g)$ from $\rho(z)$ and $g(z)$,
and defining a `wave-generating' pressure by subtracting off the
integral of this quantity.  In the zone to the left of the interface 
$(z = z_{i+1/2})$, the modified pressure is
\begin{equation}
P'_i(z) = P_i(z) - \int_{z_{i+\frac{1}{2}}}^z {(\rho g)_i (z')}dz' \quad
z_{i-1/2} \le z \le z_{i+1/2} \enskip .
\end{equation}
In the zone to the right of the interface, the modified pressure is
\begin{equation}
P'_{i+1}(z) = P_{i+1}(z) - 
 \int_{z_{i+\frac{1}{2}}}^z {(\rho g)_{i+1} (z')}dz' \quad
z_{i+1/2} \le z \le z_{i+3/2} \enskip .
\end{equation}

We may do this because the absolute value of the pressure doesn't matter
for generating waves, and only the pressure in excess of hydrostatic
equilibrium will create motion.   However, there are other effects where
the absolute value of pressure is important --- in particular, we use
$P(z)$, not $P'(z)$, in calculating sound speeds.  We also see that, if
the original reconstructions of pressure were continuous at the interface
($P_{i}(z_{i+1/2}) = P_{i+1}(z_{i+1/2})$),
then the new reconstructions will also be.

\begin{figure}

\plotone{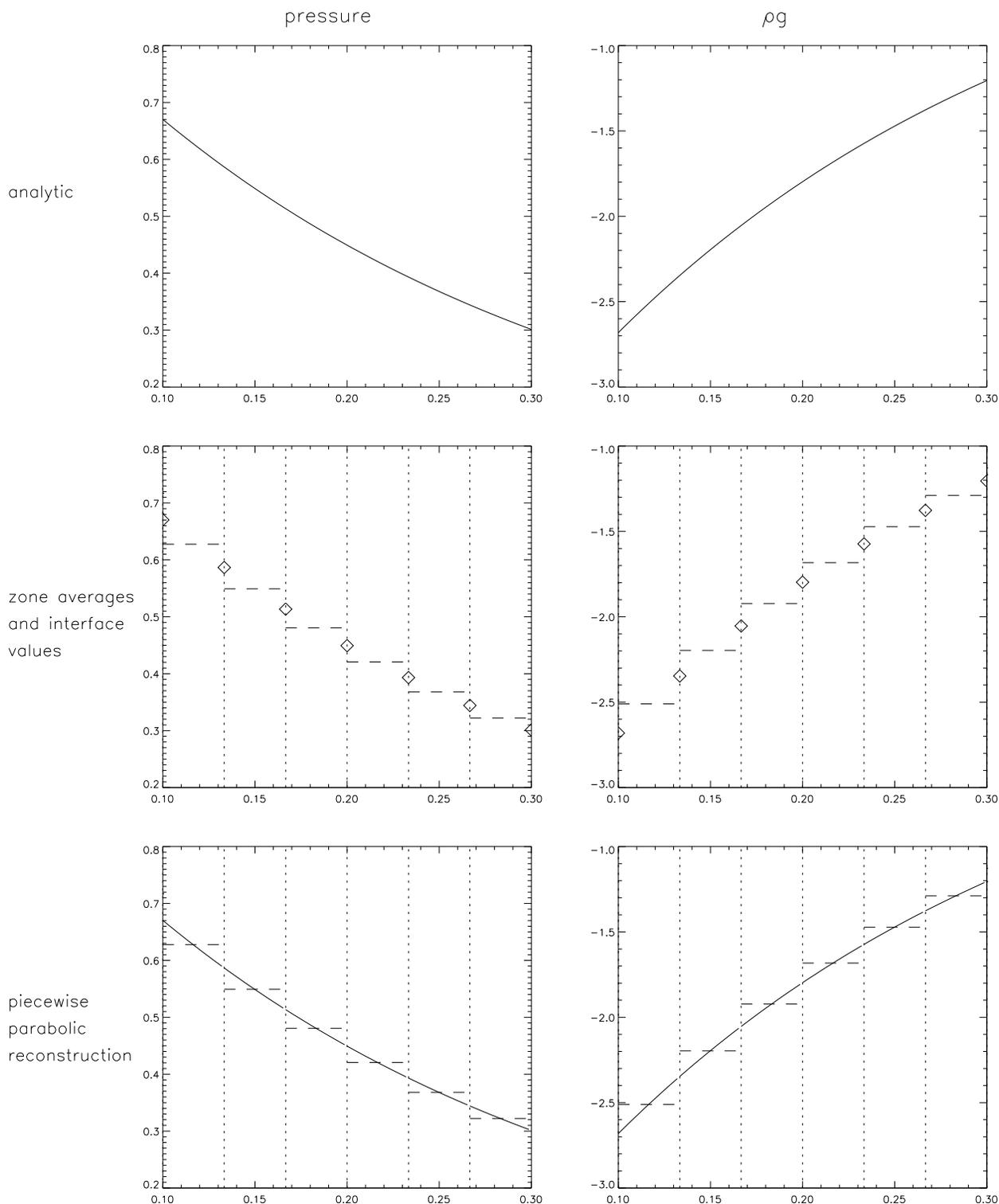}

\caption{\label{fig:modifystate} The PPM reconstruction process for a
simple exponential atmosphere, $P(z) = P_0 \exp\{-z/H\}$, where $H$ is
the scale height.  We choose $c_s = P_0 = 1$, and $g = -4$.  The left
column shows the PPM reconstruction on the pressure and the right
column shows the reconstruction on $\rho g$, which is used to derive
the wave-generating pressure.  The top row is the analytic plots of
the pressure and $\rho g$.  The second row is the zone average values
of each quantity (horizontal dashed lines) and the interface values
(diamonds), as determined using the PPM interpolants.  The zone edges
are marked with vertical dotted lines.  The bottom row shows the
piecewise-parabolic reconstructions of the original data.}
\end{figure}

\begin{figure}

\plotone{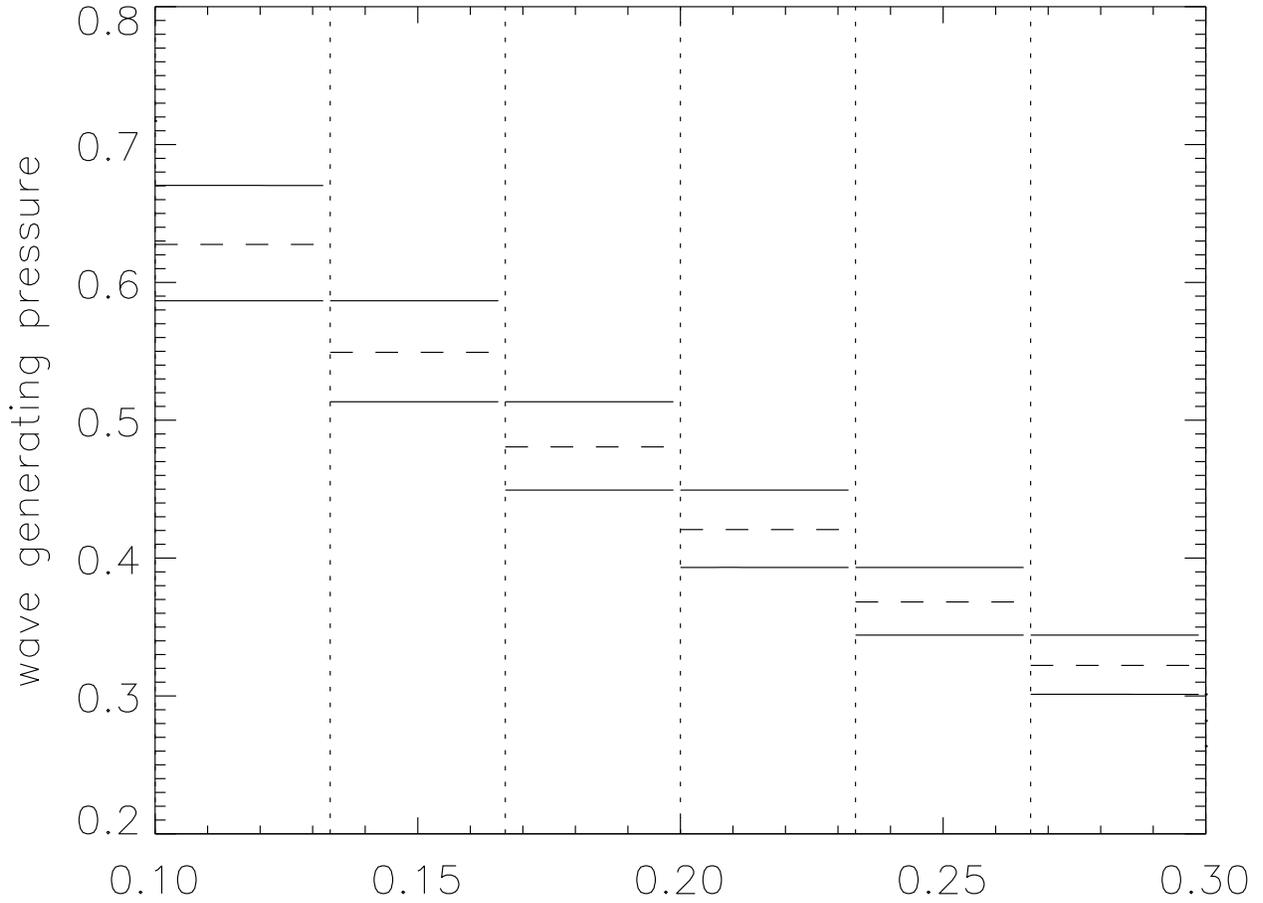}

\caption{\label{fig:modifystate2} The wave generating pressure (solid
lines) for the atmosphere shown in Figure \ref{fig:modifystate} as
computed using Eqs. (\ref{eq:ppmwavepressureright}) and
(\ref{eq:ppmwavepressureleft}).  The zone averages of the original
pressure are also shown (dashed lines).}
\end{figure}

The reconstruction of $(\rho g)$ is calculated at each point by
calculating the product of $\rho$ and $g$ and performing the same
reconstruction on this quantity as is used for quantities such as
pressure and density.  Once this (usually polynomial) reconstruction
is done, integrating it within the cell is straightforward.  The
approach is shown schematically in Figures \ref{fig:modifystate} and
\ref{fig:modifystate2}, using the method described below.

Recall that in PPM, the reconstruction is done locally by functions of the
following form:
\begin{eqnarray}
\label{eq:ppmreconstructions}
f(\alpha)  = f_{L\;i} + \alpha \left [ \Delta f_i + f_{6\;i} \left( 1 - \alpha \right) 
           \right ]; &~~~&
\alpha \equiv \frac{z - z_{i-\frac{1}{2}}}{z_{i+\frac{1}{2}} - 
                        z_{i-\frac{1}{2}}} \enskip , \\
\label{eq:ppmreconstruction2}
f(\beta)  = f_{R\;i} - \beta \left [ \Delta f_i - f_{6\;i} \left( 1 - \beta \right) 
           \right ]; &~~~&
\beta \equiv \frac{z_{i+\frac{1}{2}} - z}{z_{i+\frac{1}{2}} - 
                        z_{i-\frac{1}{2}}} \enskip ,
\end{eqnarray}
where $f_{L\;i}$, $f_{R\;i}$, $\Delta f_i$, and $f_{6\;i}$, in the
notation of \cite{flash}, are the coefficients of the reconstruction
polynomial through zone $i$ in the normalized coordinates $\alpha$ or
$\beta$.  The coefficients are chosen by the PPM algorithm to
reconstruct the data consistent with other constraints such as
monotonicity and shock capturing.  The first form of the reconstruction
is designed to be convenient when integrating through the zone from the
left (\emph{e.g.}, when the zone is on the right of the interface), and
the second when integrating from the right (\emph{e.g.}, when the zone
is on the left of the interface).

Given the above reconstructions, the integral of $(\rho g)$ in cell to
the right of the interface is (using Eq. \ref{eq:ppmreconstructions}):
\begin{eqnarray}
\int_{z_{i+\frac{1}{2}}}^z {(\rho g)_{i+1} dz} & = & \delta z_{i+1} \int_{\alpha=0}^{\alpha} {(\rho g)_{i+1}(\alpha) d\alpha} \\
   & = & \delta z_{i+1} \int_{\alpha=0}^{\alpha} \left [ (\rho g)_{L\;i+1} + \alpha \left ( \Delta (\rho g)_{i+1} + (\rho g)_{6\;i+1} \left( 1 - \alpha \right) \right ) \right ] d\alpha \\
   & = & \delta z_{i+1} \left [ - \frac{1}{3} (\rho g)_{6\;i+1} \alpha^3 + \frac{1}{2} \left ( \Delta (\rho g)_{i+1} + (\rho g)_{6\;i+1} \right ) \alpha^2 + (\rho g)_{L\,i+1} \alpha \right ]  \\
\end{eqnarray}
and in the cell to the left of the interface (using
Eq. \ref{eq:ppmreconstruction2}),
\begin{eqnarray}
\int_{z_{i+\frac{1}{2}}}^z {(\rho g)_i dz} & = & -\delta z_i \int_{\beta=0}^{\beta} {(\rho g)_i (\beta) d\beta} \\
   & = & -\delta z_i \int_{\beta=0}^{\beta} \left [ (\rho g)_{R\;i} - \beta \left ( \Delta (\rho g)_i - (\rho g)_{6\;i} \left( 1 - \beta \right) \right ) \right ] d\beta \\
   & = & -\delta z_i \left [ - \frac{1}{3} (\rho g)_{6\;i} \beta^3 + \frac{1}{2} \left ( -\Delta (\rho g)_i + (\rho g)_{6\;i} \right ) \beta^2 + (\rho g)_{R\,i} \beta \right ]  \enskip .
\end{eqnarray}
Note that, by construction, there are no terms in the integral constant
in $\alpha$ or $\beta$.

\begin{figure}

\plotone{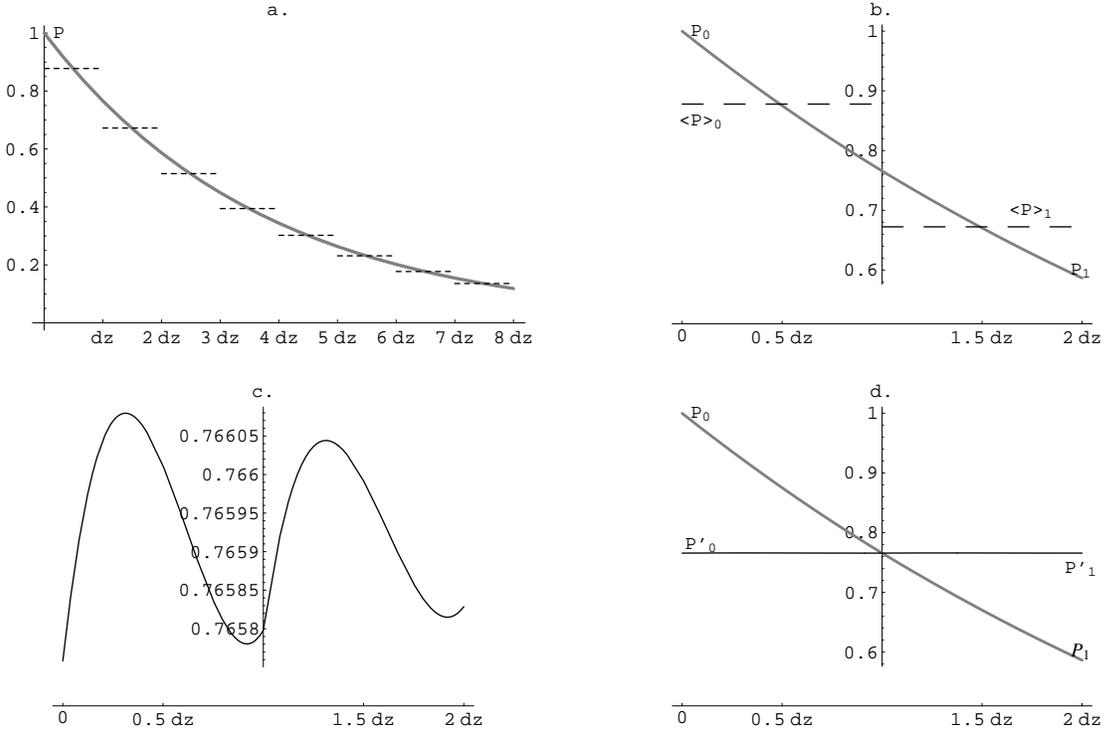}

\caption{\label{fig:demoPPMreconstruct} 
	Another, `close-up' view of the process of modifying the 
        pressure used in calculating the states to the Riemann problem.
	Shown in (a) is an exponential atmosphere, $P = P_0 \exp
	\{-z/H\}$, with $P_0 = c_s = 1$, $g =- 4$, and $\delta z = 1/15$,
	with both the PPM reconstruction as in Eq. \ref{eq:ppmreconstructions}
	(thick line) and cell-averaged values (dashed).  In (b) is
	shown only the averages and reconstructions --- as given by 
        Eqs. \ref{eq:ppmpressureright} and \ref{eq:ppmpressureleft} ---
        of the cell to the left and right of the interface at $z = \delta z$.
        Plotted in (c) is the wave-generating pressure on either side
	of the interface, as in Eqs. \ref{eq:ppmwavepressureright} and
	\ref{eq:ppmwavepressureleft}.  As can be seen, this modified
	pressure field for calculating the left and right states is
	essentially flat to the (quadratic) order of the original
	reconstruction; here, the third-order term is dominant.  In (d)
	is the wave-generating pressure and the full pressure plotted
	together.
	}
\end{figure}

The reconstructed pressure in zones $i$ and $i+1$ when integrating from
the right and from the left of the interface at $z_{i+1/2}$ is
\begin{eqnarray}
\label{eq:ppmpressureright}
P_{i}(\beta) & = &  P_{R\,i} - \beta \left [ \Delta P_{i} - P_{6\;i} 
                       \left( 1 - \beta \right) \right ] \enskip ,\\
\label{eq:ppmpressureleft}
P_{i+1}(\alpha) & = &  P_{L\,i+1} + \alpha \left [ \Delta P_{i+1} + P_{6\;i+1} 
                       \left( 1 - \alpha \right) \right ]  \enskip .
\end{eqnarray}
The new reconstructions of the pressure used in the modified states become
\begin{eqnarray}
\label{eq:ppmwavepressureright}
P'_{i}(\beta) & = &  P_{R\,i} - \beta \left [ \Delta P'_{i} - P'^{-}_{6\;i} 
                       \left( 1 - \beta \right) \right ] \enskip - 
                       \beta^3 P'_{\mathrm{cubic}\;i} \enskip , \\
\label{eq:ppmwavepressureleft}
P'_{i+1}(\alpha) & = &  P_{L\,i+1} + \alpha \left [ \Delta P'_{i+1} + 
                        P'^{+}_{6\;i+1} 
                       \left( 1 - \alpha \right) \right ] \enskip + 
                       \alpha^3 P'_{\mathrm{cubic}\;i \enskip .}
\end{eqnarray}
where
\begin{eqnarray}
\Delta P'_{i} & = & \Delta P_i - \Delta z_i 
                    \left ( (\rho g)_{L\,i} + \frac{1}{2} \left ( \Delta (\rho g)_{i} + (\rho g)_{6\;i} \right ) \right ) \enskip , \\
P'^{\pm}_{6\;i} & = & P_{6\;i} + \frac{1}{2} \delta z_i 
                    \left ( \Delta (\rho g)_i \pm (\rho g)_{6\;i} \right ) \enskip , \\
P'_{\mathrm{cubic}~i} & = & \frac{1}{3} \delta z_i (\rho g)_{6\;i} \enskip .
\end{eqnarray}
A diagram showing how this works at an interface is shown
in Figure \ref{fig:demoPPMreconstruct}.

Since these reconstructions are used to create the states for input
into the Riemann problem, the effect of gravitational acceleration is
incorporated directly into the hydrodynamics.  Thus, its effects need
not be added in later in the construction of the input states to the
Riemann solver; in particular, the terms corresponding to those
involving gravity in $\beta^{\pm}$ in Eq. (3.7) of \cite{ppm} are
removed.  It is still necessary however to add the gravitational terms
to the momentum and energy when doing the zone updates, as specified
in Eq. (3.8).  For example, the momentum is updated according to
\begin{equation}
\avgsub{\rho}{i}^{n+1} \avgsub{v}{i}^{n+1} =
\avgsub{\rho}{i}^{n} \avgsub{v}{i}^{n} + \frac{\Delta t}{\Delta V_i} 
 \left ( F_{i-1/2} - F_{i+1/2} \right ) 
+ \Delta t \left ( \avgsub{\rho}{i}^n \avgsub{g}{i}^n +
            \avgsub{\rho}{i}^{n+1} \avgsub{g}{i}^{n+1} \right ) \enskip .
\end{equation}
where $F_{i+1/2}$ is the flux through the $z_{i+1/2}$ interface.

\begin{figure}

\plotone{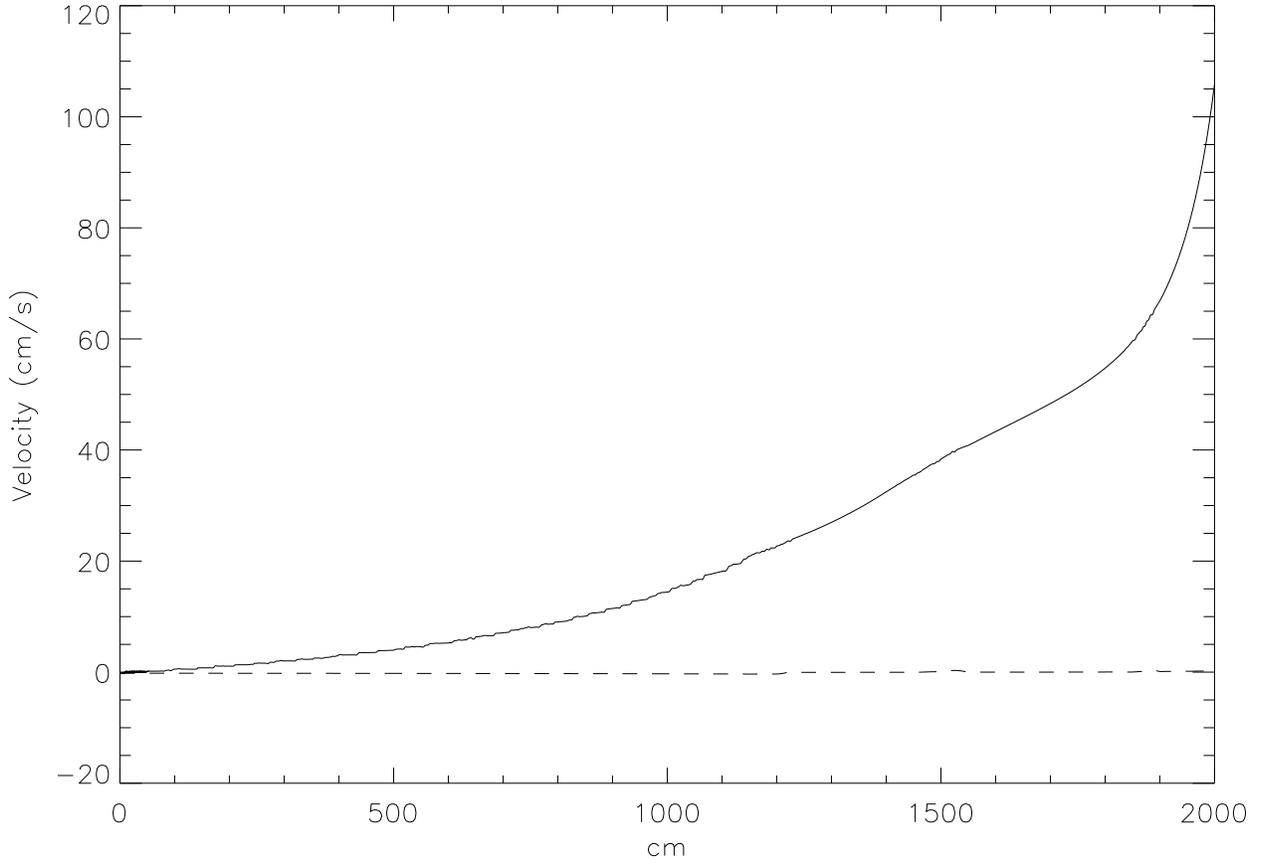}

\caption{\label{fig:hse_vs_ppm} Same as Figure \ref{fig:lvq_vs_godu},
         but using PPM (solid) and our Modified-States PPM (dashed).
	 The maximum velocity in the corrected case is 
         $0.35~{\mathrm{cm\,s^{-1}}}$.   The difference in magnitude between
         the velocities in the two figures is due to the difference in the
         accuracy of the hydrodynamic solvers (Godunov vs. PPM), not the
         result of the method used to deal with the source-term. }
\end{figure}

The results of closer coupling of the gravity to the hydrodynamics is
shown in Figures \ref{fig:lvq_vs_godu} and \ref{fig:hse_vs_ppm}.  In
both figures, the results are from the simulation of an isothermal
atmosphere, discussed in \S\ref{gamma_model}, with a pressure scale
height at the base of $1036 {\ \mathrm{cm}}$ and sound speed at the base
of $4.4 \times 10^8 {\ \mathrm{cm\,s^{-1}}}$.  In Figure
\ref{fig:lvq_vs_godu} we show the effects of using LeVeque's method in
a Godunov solver, and in Figure \ref{fig:hse_vs_ppm} we show the
results of using the above-described modified states in PPM.  In both
cases, we can see a corresponding increase in accuracy of the
solution.

Note that neither the two figures, nor the two methods of dealing with
the source-terms, can be directly compared, as the first figure
demonstrates results with a very low order Godunov method, and the
second from using higher-order PPM.   The figures demonstrate only that
our source-term correction to our hydrodynamic solver gives a similar
improvement in results as other approaches in the literature work for
other solvers.

\section{Initial Models}
To better understand the effects of the initialization, boundary
conditions, and hydrodynamics, we consider three different initial
models.  We choose the conditions of all of the initial models to
roughly agree---a base density of a few $\times\ 10^6 \gcc$ and a base
temperature of $\sim 10^7$ K, with a solar-like composition.  We use a
constant gravitational acceleration, $g = -1.8676 \times 10^{14}
\mathrm{\ cm\ s^{-2}}$. These conditions are relevant to an accreted
atmosphere on a 1.4 solar mass, $R = 10^6 \mathrm{\ cm}$ neutron
star.  The accretion phase is computed using a one-dimensional
implicit hydrodynamics code for many dynamical times, and the
subsequent burning front propagation is then followed using 
a multidimensional hydrodynamics code.

Once the initial model is set up, it will be mapped onto the computational
grid.  The PPM algorithm we use carries the cell averaged value of each
variable in each zone.  For an initial model that is not analytic,
but rather a series of points and corresponding values, we will need
to map onto the grid with the understanding that the values on the new
grid are treated as zone averages.  We will consider several different
initial models of increasing complexity below.

\subsection{\label{gamma_model} Isothermal Ideal Gas EOS Atmosphere}

The simplest initial model one can imagine is an isothermal atmosphere
in hydrostatic equilibrium, whose pressure is obtained via the ideal gas law,
\begin{equation}
\label{eq:hse_ideal}
\frac{d P}{d z} = - |g| \rho(z) ; 
\quad P = \frac{k_B \rho T}{\bar{A} m_u} = \rho c_s^2 \enskip ,
\end{equation}
where $P$ is the pressure, $g$ is the constant gravitational acceleration,
$\rho$ is the density, $k_B$ is the Boltzmann constant, $T$ is the
temperature, $\bar{A}$ is the average atomic mass, $m_u$ is the
nucleon mass, and $c_s$ is the isothermal sound speed.
Although this EOS is not relevant to a neutron star atmosphere, we
will use it to examine some of the properties of hydrostatic
equilibrium in a Godunov type code.

Eq. (\ref{eq:hse_ideal}) can be integrated analytically to yield an
exponential atmosphere,
\begin{equation}
 \rho(z) = \rho_0 \cdot \exp\{- |g| z / c_s^2\} = \rho_0 \cdot \exp \{-z/H\} \enskip ,
\end{equation}
where $\rho_0$ is the base density, and
\begin{equation}
 P(z) = \rho_0 c_s^2 \exp\{-z/H\} = P_0 \exp\{-z/H\} \enskip .
\end{equation}
The quantity $H = c_s^2/|g|$, the scale height of the atmosphere, 
will play an important role in determining the resolution necessary to
hold any of the initial models steady.  If $H < dz$, where $dz$ is the
zone size, then the model will be very under-resolved, and we can
expect to have great difficulty in maintaining HSE.  

To see this, consider modeling the exponential atmosphere with a
parabola, similar to the way a second-order Godunov-type code will
model the physical quantities.
If the pressure at $z=0$ is $P_0$, and the left and right
points are $\delta$ on either side with cell-averaged pressures 
$\avgsub{P}{-1}$ and $\avgsub{P}{+1}$ (from Eq. \ref{eq:quadinterp}), 
we can construct the quadratic 
that has the same cell-averaged quantities as the exponential;
in particular, the derivative of pressure over the cell containing $z=0$ is
\begin{equation}
\avgsub{\frac{dP}{dz}}{0} = \frac{\avgsub{P}{+1} - \avgsub{P}{-1}}{2 \delta} \enskip .
\end{equation}
The cell-averaged pressures on the left and right will be 
\begin{equation}
\avgsub{P}{-1} = P_0 \frac{H}{\delta}\left( \exp\{-\delta/(2H)\} - \exp\{-3 \delta/(2H)\}\right )
\end{equation}
and
\begin{equation}
\avgsub{P}{+1} = P_0 \frac{H}{\delta}\left( \exp\{3 \delta/(2H)\} - \exp\{\delta/(2H)\}\right ) \enskip , 
\end{equation}
respectively.  (This comes from integrating
the exponential pressure, $P(z)$ over the left and right
cells, which range from $(-3\delta/2,-\delta/2)$ and
$(\delta/2,3\delta/2)$ respectively.)  With the above equation and
$P_0 = \rho_0 g H$, and defining $x \equiv \delta/H$, the
acceleration (in units of $g$) at $z=0$ is
\begin{equation}
\frac{1}{\rho_0 g} \avgsub{\frac{dP}{dz}}{0} = -\frac{1}{x^2} \left ( \cosh \left(\frac{x}{2} \right) - \cosh \left(\frac{3x}{2} \right) \right ) \enskip ,
\end{equation}
which, to second order in $x$, is $-1 - (5/24) x^2$.  The first term
of this acceleration is simply the gravitational acceleration, leaving
an error in acceleration of magnitude $\frac{5}{24} x^2$.  Per CFL
timestep, this corresponds to a spurious velocity, expressed as a Mach
number, of
\begin{equation}
{\cal M} \sim \frac{5}{24} x^3 \enskip .
\end{equation}

Thus, if one wishes to consider very subsonic flows ${\cal M} \sim
10^{-3}$ moving through a stratified atmosphere, one needs at
\emph{least} 6 points per pressure scale height, or else one will get
velocities of greater magnitude every timestep.  Since the velocity
errors are cumulative, one would realistically need much more than
that.

\subsection{\label{realeos_atm} Realistic EOS Atmosphere}

The next complication we can imagine adding to our initial model is
using a more realistic equation of state.  This more generalized EOS
prevents us from integrating the model analytically as above, but we
can still compute the model on an arbitrary grid by simply differencing the
equation of HSE (Eq.~\ref{eq:hse_full}).  To recover cell averaged
quantities, we can subsample each zone and compute the average.

We use a Helmholtz free energy, table-based EOS for a degenerate
electron gas with perfect gas ions and radiation pressure included
with FLASH \citep{eos} for this model.  The material at the base of
this model is partially degenerate, and this EOS accurately describes
the thermodynamics in our neutron star atmosphere.  To complete this
model, we again assume that our atmosphere is isothermal.  The
atmospheric structure is completely determined by the choice of
temperature (we use $3 \times 10^7$ K), the composition (we assume 0.7
$^1$H and 0.3 $^4$He by mass), and the density at the base (we take $5
\times 10^6 \gcc$).  Because of the complexity of the EOS, despite the
fact that this is an isothermal atmosphere, the scale height is not
constant.

The pressure and density profiles are computed as follows.  At each
point, the $X_i$ and $T$ are taken as given, and the equation of
hydrostatic balance,
\begin{equation} 
\label{eq:hse} \frac{d P}{d z} = g \rho \enskip ,
\end{equation} 
is numerically integrated outwards, using reconstructions to the data,
as described below.  This equation, plus the equation of state, is enough to
determine (by iteration) both the pressure and density separately.

Given the cell-averaged pressure and density data from the previous
points and the functions in \S \ref{sec:cellavg}, we can construct
functions for the pressure and density field up to and including the
current cell.  Given two such functions, one for density and one for
pressure, we can use Eq. (\ref{eq:hse}) to relate $\avgsub{\rho}{+1}$
and $\avgsub{P}{+1}$.

If we use a linear function to model the density, as in Eq.
(\ref{eq:linearinterp}), we must use a quadratic one to model the
pressure (because of the derivative), as in Eq. (\ref{eq:quadinterp}).
Using these expressions in Eq. (\ref{eq:hse}), and solving for
$\avgsub{P}{+1}$ and $\avgsub{\rho}{+1}$, we get
\begin{equation}
\label{eq:firstorderdiff}
\avgsub{P}{+1}-\avgsub{P}{0}  = \frac{g \delta}{2} \left ( \avgsub{\rho}{+1} + \avgsub{\rho}{0} \right ) \enskip ,
\end{equation}
which, with the EOS, is enough to determine $\avgsub{P}{+1}$ and
$\avgsub{\rho}{+1}$.  We will refer to this expression as our
first-order differencing.

For more accuracy, we can use a quadratic to fit the density, and a
cubic to fit the pressure.  The cubic is given by
Eq. (\ref{eq:cubicinterp}).  We then find
\begin{equation}
\label{eq:secondorderdiff}
\avgsub{P}{+1}-\avgsub{P}{0}  = \frac{g \delta}{12} \left ( 5\avgsub{\rho}{+1} + 8\avgsub{\rho}{0} 
                  - \avgsub{\rho}{-1} \right ) \enskip .
\end{equation}
We will use both of these differencing schemes in \S \ref{results} to
assess how large of a difference the choice makes.

A final difference scheme, using a quartic (Eq. \ref{eq:quarticinterp})
and a cubic, can be constructed:
\begin{equation}
\label{eq:thirdorderdiff}
\avgsub{P}{+1}-\avgsub{P}{0}  = \frac{g \delta}{12} \left ( 
     9\avgsub{\rho}{+1} + 19\avgsub{\rho}{0} - 5\avgsub{\rho}{-1} 
     + \avgsub{\rho}{-2} \right ) \enskip .
\end{equation}
but this extra accuracy is not expected to be significant for a second-order
accurate code, and numerical experiments confirm this.

\begin{figure}

\plotone{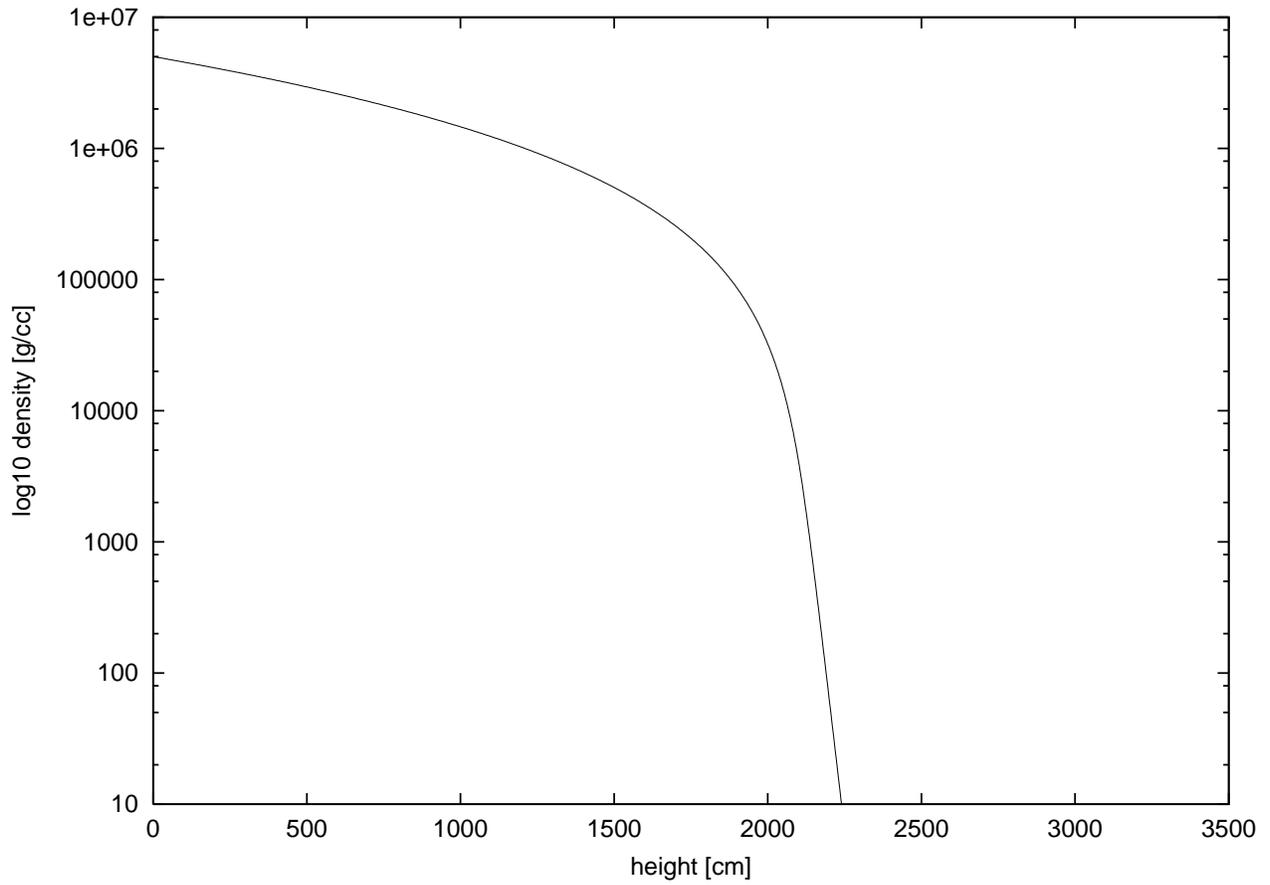}

\caption{\label{isothermal_model} The density profile of an
isothermal, uniform composition initial model with a complex EOS.
Above $2000 \ \mathrm{cm}$, the density rapidly falls.}

\end{figure}

\begin{figure}

\plotone{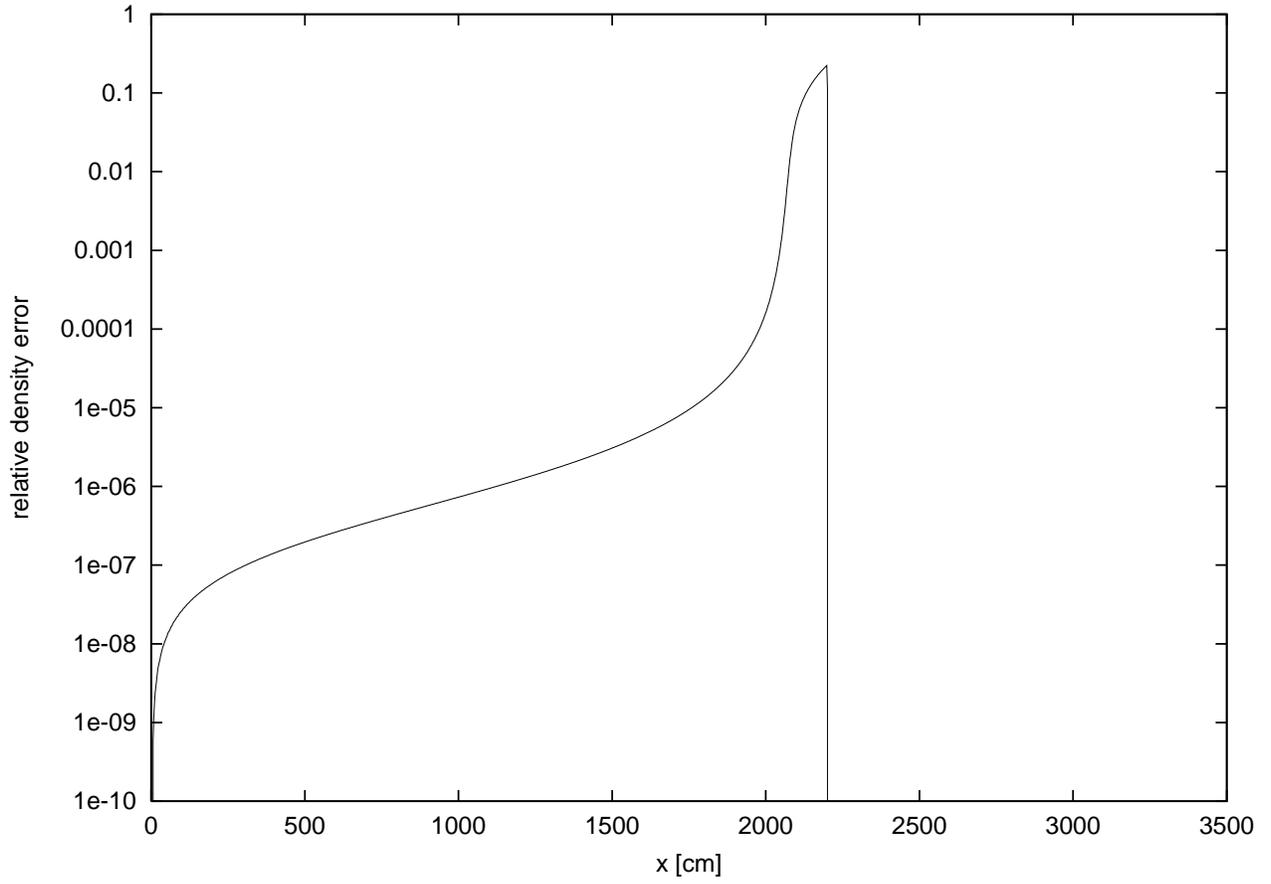}

\caption{\label{order_diff} Relative difference in the density
structure for the first-order vs. second order differencing.  The
error is greatest just before the low density cutoff.}

\end{figure}

To generate our model, we take the base density and pressure and use
either Eq. (\ref{eq:firstorderdiff}) or (\ref{eq:secondorderdiff}) to
find the pressure and density in the next zone.  We
continue this procedure, moving outward from the base, until the
density falls below a small density cutoff we impose ($10^{-5} \gcc$).
Figure \ref{isothermal_model} shows a plot of this model.  We notice
that as the density changes, the scale height also changes.  At the
top of the atmosphere, the small scale height is reflected in the
steepness of the density just before our low-density cutoff.  The
sound speed at the base of the atmosphere is $\sim 4\times 10^8 \cms$,
giving a dynamical timescale for the atmosphere of $\sim 5 \micros$.
Figure \ref{order_diff} shows the relative error in the density of the
model when using the first-order vs. second-order differencing.

\subsection{\label{stellar_evol_model} Model from a 1-d Stellar Evolution Code}

The final model we consider is one that comes directly from the
Kepler 1-d stellar evolution code \citep{kepler}.  This model was
generated by accreting H/He onto the surface of a neutron star,
gradually building up a fuel layer 17 meters thick.  Nuclear burning,
a complex equation of state, and convection were included in this
calculation.  The evolution was stopped shortly before runaway, and
the velocities in the atmosphere are small, but non-zero.
Additionally, this code solved the hydrodynamics equations in a
Lagrangian formulation, requiring a translation from mass-based zones
to an Eulerian grid.  The number of grid points in the Kepler model is
much smaller than the number of points on our Eulerian grid.

We note that in contrast to the other models described above, the
composition in this model is not uniform, but abruptly changes at the
transition from the accreted fuel layer to the underlying neutron
star, which is comprised of iron.  This in turn creates a density
jump, since the ionic component of the pressure scales as $1 /
\bar{A}$.

To import this model onto the FLASH grid, we use the following
prescription.  First, the model is re-gridded from the original
Lagrangian mesh onto a uniform, one-dimensional Eulerian grid, whose
resolution is equal to the finest spatial resolution on the
multidimensional hydrodynamic grid.  Once interpolated, we
renormalized the abundances.  Varying precision in the initial model
data file compounded by interpolation errors may result in a set of
abundances in a zone that do not sum to unity; we divide by the actual
numerical sum of the abundances to enforce the constraint of mass
fractions summing to one.  Next we take the temperature, density, and
composition as given, and update the remaining thermodynamic variables
in the zones with the FLASH EOS.  In practice, the change in the
thermodynamic variables caused by the differences in the EOS is small,
$< 1\%$.

We now want to restore hydrostatic equilibrium to this model, after
adjusting the thermodynamic variables.  The velocities are set to zero
in every zone, since we are not presently concerned with how to map a
one-dimensional velocity field into multiple dimensions.  If the model
was really from a slowly simmering phase, this is not a bad
approximation; in the model used in these calculations, for instance,
the maximum velocity was $\sim 10^{-1} \cms$.  To restore this model
into hydrostatic equilibrium, we must pick a point in the model whose
$\rho$, $T$, and $X_i$ will remain fixed, and integrate outward from
there.  We show two choices here, (i), taking the bottom-most zone as
the starting point, and (ii), taking the point just above the
composition interface as the starting point.

The differencing is performed in the same manner as described in the previous
section.  We experiment with both differencing schemes,
Eqs.\ (\ref{eq:firstorderdiff}) and (\ref{eq:secondorderdiff}).  This
differencing is continued throughout the entire model.  When
integrating toward the top of the atmosphere, we stop putting the
model into hydrostatic equilibrium once the density becomes so small as
to be no longer dynamically important.  We use a cutoff value for the
density of $1 \times 10^{-5} \gcc$, and affectionately refer to the
material above this as the ``fluff''.  This cutoff is needed since we
cannot continue the HSE profile to arbitrary heights, as the densities
would quickly underflow.

We note that instead of adjusting the density along with the pressure,
one could adjust the temperature.  For cases where the atmosphere is
degenerate (like our present case), this can be problematic due to
the insensitivity of the pressure on the temperature.  We do not
explore this approach in the present paper.

\begin{figure}

\epsscale{0.65}
\plotone{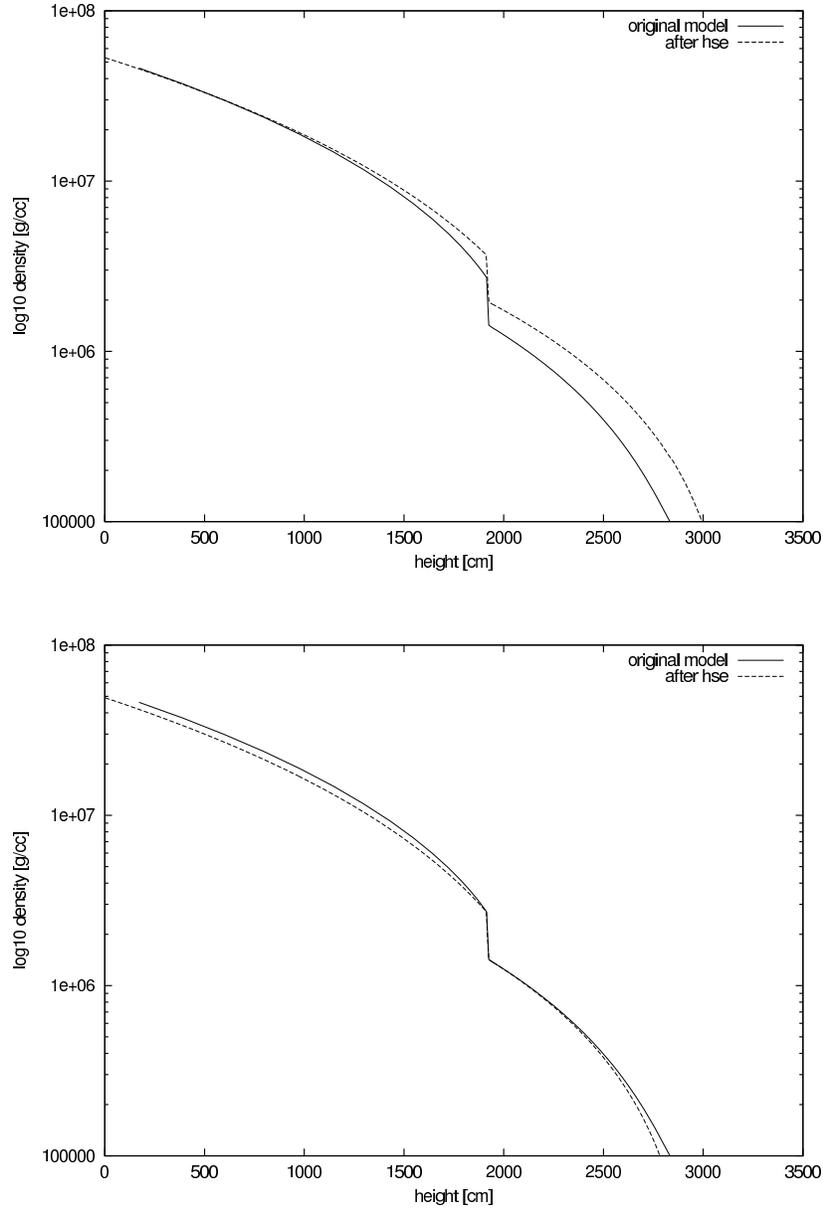}

\caption{\label{kepler_model} Results of taking a 1-d initial model
from the Kepler implicit stellar hydrodynamics code and putting it into HSE
with the new EOS.  We note the jump in density in this model, owing to
the abrupt change in composition from the underlying neutron star to
the accreted fuel layer.  In the top panel, we took the base of the
initial model as the reference point.  In the lower panel, we took the
base of the fuel layer in the initial model as the reference point.}

\end{figure}

Figure \ref{kepler_model} shows the results of applying this procedure
to our Kepler initial model.  The two panels differ in the choice of
the reference point used when differencing the model into HSE.  We
note that in the top panel, where we chose the base of the model to
begin the integration, the errors compound greatly as we integrate
outward, especially where we have to integrate through a material
discontinuity, which is poorly modeled by a low-order polynomial.  At
the base of the fuel layer, there is a significant difference in the
density.  The agreement is much more uniform in the lower panel, since
the reference point (the base of the fuel layer) is near the center of
the model. We use the base of the fuel layer as the reference point in
all Kepler-model simulations presented here.

\section{Boundary Conditions}

The stability of the model atmosphere can be very sensitive to the
choice of boundary conditions.  In these highly-subsonic simulations,
the hydrodynamic equations are essentially elliptic, so that
boundary conditions matter as much as the initial conditions inside the
computational domain.  We investigate many different boundary conditions
at the lower boundary, and two different boundary conditions at the
upper boundary in this study.

In FLASH, like most finite-volume hydrodynamics codes, the boundary
conditions are implemented in fictitious zones outside of the physical
domain called guardcells or ghostcells.  In order to allow for
refinement and parallelization of the code, the computational domain
is broken into multiple sub-domains, or blocks.  Each block is
surrounded by a perimeter of guardcells that hold the data from
neighboring blocks, or, if at a physical boundary of the computational
domain, are filled with the proper boundary condition.

The problem of generating good hydrostatic boundary conditions is
closely related to that of finding good `outflow' or non-reflecting
boundary conditions (\citealt{outflow}), which remains an area of
research.  In both situations, to create a desired flow condition, one
has to set up corresponding physical fluid conditions, essentially
inverting the Riemann problem.  Further compounding the difficulty is
that the problem must be solved in a way consistent with one's
hydrodynamic solver.  Below are several boundary conditions, both
commonly-used in the astrophysical literature and novel, which are
approximations of solutions to the inverse problem.

\subsection{\label{lower_bc} Lower Boundary}

The lower boundary must support the weight of the atmosphere above it,
while still allowing for dynamics.  We consider two classes of
boundary conditions --- a standard reflecting boundary, and a
hydrostatic boundary, which provides pressure support to the material
above while still allowing flow through the boundary.

\subsubsection{Reflecting}

The simplest lower boundary we can use that will support the weight of
the material above it is a reflecting boundary.  This is one of the
most commonly used boundary conditions when evolving a hydrostatic
atmosphere.  Absent any gravity, this boundary condition simply
reverses the sign of the normal velocities in the boundary region, and
gives all other variables a zero-gradient.  The result is that there
is no flow through the boundary.  Any wave that hits the boundary is
reflected back into the interior.  This boundary condition is
effective at restricting flow through the boundary, but it will not
allow sound waves to leave the grid as the initial model relaxes.

In the presence of gravity, the traditional reflecting velocity 
boundary condition of `flipping' the velocity in the direction transverse
to the boundary, 
\begin{equation}
\label{eq:simple_reflect}
v_z(z - z_0) = -v_z(z_0 - z) \enskip ,
\end{equation}
(where $z_0$ is the location of the physical boundary) will not work,
because there is an acceleration term due to gravity.  Better is
\begin{equation}
v_z(z - z_0) = -v_z(z_0 - z) + g \delta t \enskip ,
\end{equation}
which takes into account the acceleration performed at
the hydro step when integrating the equations.  In a finite
volume-code it is easy, in addition, to enforce the desired
`no-penetration' condition exactly by setting the flux across the
interface defining the boundary to exactly zero.  This is done in the
reflecting boundary condition results below.

\subsubsection{Hydrostatic}

\begin{figure}

\epsscale{0.95}
\plotone{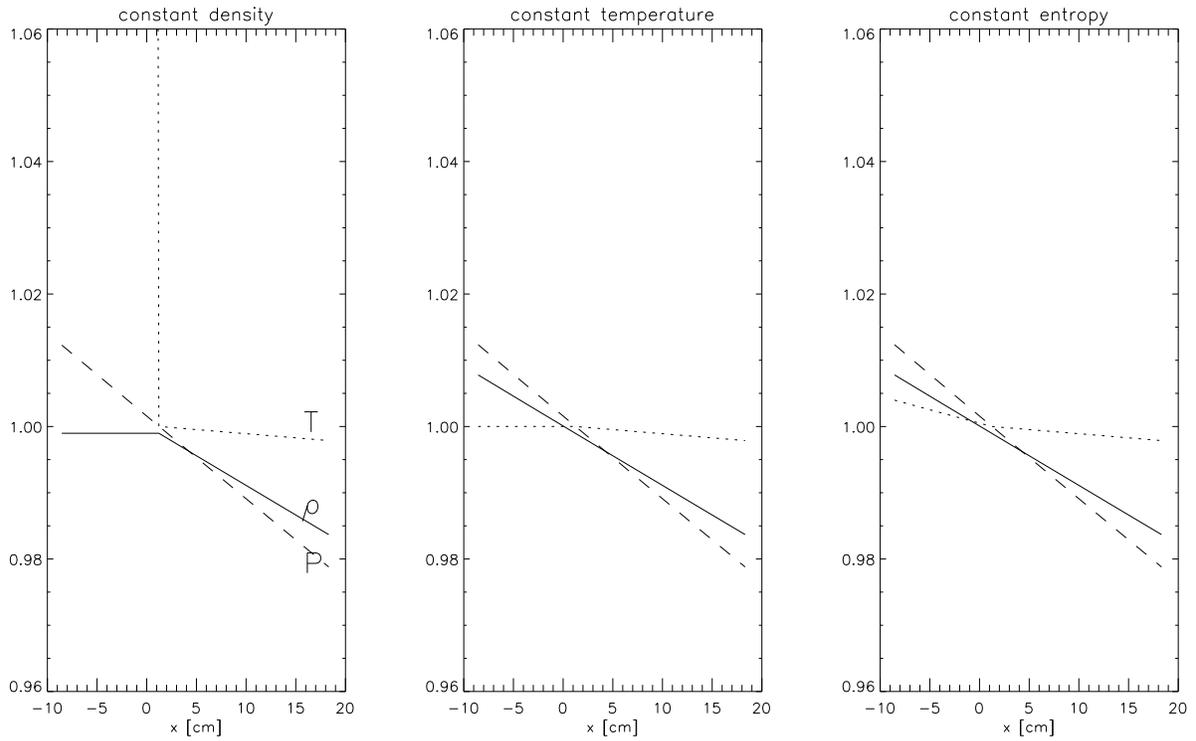}

\caption{\label{fig:bc_guardcells} Comparison of the three different
hydrostatic boundary conditions, showing constant density (left),
constant temperature (center), and constant entropy (right).
Pressure, density, and temperature are plotted, scaled to their base
values ($x$=0).  Negative values of $x$ are the guardcells.  The
temperature in the constant density case grows very rapidly in the
guardcells, due to the degeneracy of the gas.}

\end{figure}

The reflecting boundary is artificial in the sense that it does not
allow waves to flow off the grid.  An alternative is to use a boundary
condition that understands hydrostatic equilibrium.  The basic
strategy is to provide pressure support to the material above the
boundary by solving the equation of hydrostatic equilibrium in the
boundary region.  This is done by simple differencing as described
above.  When filling the guardcells, an additional constraint on the
material is needed, since the HSE equation does not uniquely determine
the boundary states given the states above.  We assume that the
composition and either density, temperature, or entropy are constant.
This is not a complete set of possible constraints, but is enough to
illustrate the effect this can have on the evolution of an atmosphere.
Any constraint should be motivated by the physics of the model under
study.  Figure \ref{fig:bc_guardcells} shows the density, temperature,
and pressure at the base of the Kepler model (including guardcells)
for the different choices of constant variable.  We see that they all
yield roughly the same pressure profile, since the pressure at any
point in the atmosphere is determined by the weight of the material
above it:
\begin{equation}
  P = \int \rho g \ dz = g \sigma \enskip ,
\end{equation}
where $\sigma$ is the column density.  Because of the degeneracy of
the gas, we see that the temperature must rise dramatically in the
guardcells in the constant density case in order to provide the needed
pressure support.

It is important to note that the best boundary condition may be
problem dependent.  Differences in the EOS or the physics of the
atmosphere will affect the choice of the constraint.

\subsubsubsection{constant $\rho$, first order}

The easiest hydrostatic boundary to implement uses the simple first-order
differencing of the hydrostatic equilibrium equation
(Eq. \ref{eq:firstorderdiff}).  In \S \ref{realeos_atm}, we derived
this to find the pressure in the zone immediately above the current
zone.  At the lower boundary, we need to find the pressure in the zone
below the last interior zone, given the pressure and density in that
zone.  Thus, our difference equation becomes:
\begin{equation}
\label{eq:bc_first_order}
\avgsub{P}{0}-\avgsub{P}{+1}  = -\frac{g \delta}{2} \left ( \avgsub{\rho}{0} + \avgsub{\rho}{+1} \right ) \enskip .
\end{equation}
The right hand side requires the density in the $0^{th}$ zone.  In
this case, we assume that the density is constant (zero-gradient) in
the guardcells, initializing all of them with the value of the density
in the first interior zone.  Therefore in order to satisfy the EOS,
the temperature will rise in the guardcells.

All other variables are also given a zero-gradient, except the
velocity.  We use three different methods for dealing with the
velocity.  The first is to give it a zero gradient, so the velocity in
the guardcells is constant, and equal in magnitude and direction to
the velocity in the last interior zone.  We call this the ``outflow''
condition.  The next method is to perform the outflow method only if
the velocity is leaving the grid.  If the sign of the velocity in the
first interior zone is positive, indicating that material is flowing
onto the grid, we set the velocity in the guardcells to 0.  We call
this the ``diode'' boundary condition.  The final method is to simply
reflect the velocity, as indicated by Eq. (\ref{eq:simple_reflect}).

\subsubsubsection{constant T, first order}

A slight variation on the above is to give the temperature a
zero-gradient in the guardcells, initializing it to the value in the
first interior zone.  Then the density will need to increase (as given
by the EOS) in order to give the pressure demanded by Eq
(\ref{eq:bc_first_order}).  We also experiment with the three different 
velocity methods described above.

\subsubsubsection{constant entropy, first order}

A more realistic boundary condition for a stellar model is to make the
temperature/density profile isentropic and in HSE.  For
an atmosphere in hydrostatic equilibrium the adiabatic temperature
gradient is
\begin{equation}
\label{eq:ad_T}
\left ( \frac{d T}{d z} \right )_{ad} = \frac{\delta_P g}{c_P} \enskip ,
\end{equation}
(see, for example, \citealt{lantz_fan}) where
\begin{equation}
\delta_P = - \left ( \frac{\partial \ln \rho}{\partial \ln T} \right ) \enskip ,
\end{equation}
and $c_p$ is the specific heat at constant pressure.
Together with the equation of hydrostatic equilibrium, this specifies
the conditions in the guardcells.  We still have to assume a function
for the velocities, and we use the same choices as above.

To implement this boundary condition, we use a first order
differencing for the HSE equation (Eq. \ref{eq:bc_first_order}), and a
simple first order differencing of Eq. (\ref{eq:ad_T}).  
These two expressions are solved simultaneously along with the EOS and
iterated until we obtain convergence of the pressure and density.

\subsubsubsection{constant $\rho$, second order}

We can use a variation of Eq. (\ref{eq:secondorderdiff}) to perform the
differencing in the guardcells.  This is a higher order differencing
than that used in the above boundary conditions, and should give more
accurate results.  Again, we are interested in using the information
in the zones just above the lower edge of the computational domain to
fill the guardcells below it, so the differencing is in the opposite
direction as before:
\begin{equation}
\label{eq:bc_second_order}
\avgsub{P}{0}-\avgsub{P}{+1}  = -\frac{g \delta}{12} \left ( 5\avgsub{\rho}{0} + 8\avgsub{\rho}{+1} 
                  - \avgsub{\rho}{+2} \right ).
\end{equation}
Again we give all variables except velocity a zero-gradient, and
retain our three choices for dealing with the velocity as described
above.  In this case we keep the density constant in the guardcells
and the pressure given by Eq (\ref{eq:bc_second_order}) as a
constraint on our EOS to give the temperature in the guardcells.  Thus
the temperature will rise with depth in our boundary.

\subsubsubsection{constant T, second order}

This case is the same as above, but we take the temperature as constant in
the boundary zones and solve Eq (\ref{eq:bc_second_order}) in tandem
with the EOS to find both $\avgsub{P}{0}$ and $\avgsub{\rho}{0}$
simultaneously.  Again, we allow for all three different velocity
choices.

\subsubsubsection{constant entropy, second order}

This case is another implementation of the constant entropy boundary
condition defined above, but we use the second order differencing for
the HSE equation (Eq. \ref{eq:bc_second_order}), and keep the first
order differencing for the temperature expression.  The pressure and
density are the two variables that are involved in the dynamics, so
this is why these are treated as second order.

\subsection{Upper Boundary}

The choice of the upper boundary is also important.  Further out in
the hydrostatic envelope, eventually the density will reach a very
small value, such that it can no longer be represented in IEEE
floating point arithmetic without underflowing.  Figure
\ref{isothermal_model} illustrates this effect nicely.  Above 2000 cm,
the density drops off very rapidly.  If we were to continue to follow
the density down to arbitrarily small values, the change in the height
of the atmosphere would be insignificant, to the point where it would
be less than a computational zone.  There are two major paths one may
take to get around this.  First, one may only put a portion of the
atmosphere on the grid, leaving off the top few scale heights.  In
this case, a hydrostatic boundary condition is required at the top,
preferably one that makes the same assumptions as the lower boundary.
The difficulty with this method is that, if in the dynamics that
follow in a simulation the envelope heats up, the scale height will
get larger, and more and more of the envelope will be lost through the
top boundary.  This can be overcome partially by employing an
expanding grid, but that case is not considered here.  Once material
leaves the top of the grid, it is lost forever, and would be unable to
further contribute to the dynamics of the simulation.

An alternative is to follow the hydrostatic structure down to a cutoff
value of density (something large enough still so it will not
underflow), and then apply a uniform density above this point in the
atmosphere.  This creates the `fluff' described in \S
\ref{stellar_evol_model}.  The top boundary can then be a zero
gradient, or impose some inflow characteristic of accretion.  Since we
do not place this material in hydrostatic equilibrium, it will
fall under the influence of gravity.  However, its mass is so small
that it will have very little dynamical impact.  Leaving a buffer of
the low-density material above the hydrostatic envelope allows the
envelope to expand in the computational domain.  In particular,
expansion that is non-uniform can more easily be handled in this case
than with an expanding Eulerian grid.  This choice of boundary
condition allows the exploration of surface features, such as waves,
that are created as localized heating causes some regions of the
envelope to expand before others.  We choose this avenue for most
of the simulations presented in this paper,  and study the effect
of this choice in \S \ref{result:upper_bc}.

\section{Results}
\label{results}

We ran simulations using both the isothermal, realistic EOS
atmosphere, and the Kepler generated initial atmosphere.  Both initial
models were run many times, varying the resolution, grid type,
boundary conditions, and hydrodynamics.  Unless otherwise noted, all
simulations use the standard PPM algorithm.  Table \ref{parameters}
summarizes the different parameters we explore.  All calculations were
performed in one dimension, for computational efficiency.  We examine
the effect of each of these parameters in turn below.

\subsection{Effect of Boundary Conditions}

\subsubsection{\label{result:lower_bc} Lower BC}

As discussed in \S \ref{lower_bc}, there are a large number of
variations of the hydrostatic boundary condition.  We consider the
effect of the boundary conditions first, since, as we will see, the
choice of boundary condition has a large effect.  To examine the
influence of all these differences, we ran each of our initial models
on a uniform grid with the standard PPM algorithm, varying the
parameters in the boundary conditions.  All of these calculations have
a zone width of 2.4 cm and were run for $250\micros$ ($\sim$ 20
dynamical times).  The results are summarized in Figures
\ref{fig:iso_bc_plots} and \ref{fig:bc_plots}.

A total of 10 cases are examined here for both the Kepler initial
model and the isothermal atmosphere initial model, showing both the
density and velocity profiles at the end of the calculation
($250\micros$), the total momentum as a function of time, and the
kinetic energy as a function of time.  These latter two plots allow us
to understand how the velocities are evolving with time---whether they
are relaxing down to a quiet state (as we hope), or increasing without
bound.  Reflecting and hydrostatic boundaries were used at the lower
boundary, and a fluff condition was used at the top.  We note that
this momentum includes that of the fluff, but since its density is so
low, we expect it to only make a minor contribution.  Despite our best
efforts at initializing the grid with our model in a manner consistent
with our choice of hydrodynamics algorithm, some velocities are
quickly generated as we evolve our models in time.  All simulations
were initialized using the second order differencing and the second
order variants of the different boundary conditions were used.
Ideally, the velocities would be small in magnitude and diminish
quickly as the model relaxes.  We will look at the effect of the
initialization method in a latter section.

We note a number of things right away---the hydrostatic boundary
conditions that keep the temperature or entropy constant do a far
better job than those that keep the density constant.  In Figures
\ref{fig:bc_plots}d-e, we note that the Kepler atmosphere is falling off the
bottom of the grid and large negative velocities are dominating in the
atmosphere.  In the case of hydrostatic with constant density, but
reflecting velocities (Figure \ref{fig:bc_plots}f), we note that the
atmosphere is held in the box, due to the reflection of the
velocities.  We refer back to Figure \ref{fig:bc_guardcells}---the
extreme rise in temperature in the guardcells needed to support the
weight of the material above the boundaries is the likely culprit
here.  This situation is much more severe for the isothermal
atmosphere case (Figure \ref{fig:iso_bc_plots}d-e), as the model quickly
($< 50\micros$) falls through the bottom of the domain.

The hydrostatic conditions that use a simple zero-gradient/outflow
condition on the velocities for the constant temperature case (Figures
\ref{fig:iso_bc_plots}a and \ref{fig:bc_plots}a) show the momentum
monotonically increasing with time, while the constant entropy case
(Figures \ref{fig:iso_bc_plots}g and \ref{fig:bc_plots}g) show the
momentum monotonically decreasing with time.  The velocity is very
flat in the hydrostatic atmosphere, with the exception of a `hiccup'
at the composition interface of the Kepler model.  However, this
velocities steadily increases in magnitude with time, making this
condition ill-suited for long simulations.

Any condition (hydrostatic or not) that reflects the velocities at the
lower boundary (Figures \ref{fig:iso_bc_plots}c,f,i-j and
\ref{fig:bc_plots}c,f,i-j) shows a ringing which may be observed in
the momentum and energy plots, with a period about equal to the
dynamical timescale for the envelope.  This is due to the reflection
of the velocities, which, with the hydrostatic state, diminishes the
waves that can penetrate through the boundary.

The diode constraint coupled with the constant temperature hydrostatic
boundary condition also shows a ringing, with an amplitude (see
Figures \ref{fig:iso_bc_plots}b and \ref{fig:bc_plots}b) that is on
the order of, or less than that of the reflecting boundaries (compare
to Figures \ref{fig:iso_bc_plots}c and \ref{fig:bc_plots}c).  It also
appears to be decreasing in amplitude with time.  After an initial
transient, the diode boundary conditions give the lowest velocities in
the envelope of any of the boundary conditions tested.

We can also look at the effect of the order of the HSE differencing in
the boundary condition on the velocities in the envelope.  In all
cases, the velocities and magnitude of the momentum are smaller with
the 2nd order differencing (holding the other parameters constant).

The boundary condition that leads to the quietest velocity field,
after the initial transients die down, is the hydrostatic boundary
with constant temperature and diode velocities.  For the Kepler model,
this boundary condition gives a momentum after $250 \micros$ that is 2
orders of magnitude smaller than the reflecting boundary.  We stress
again that this finding may be problem dependent, and tests should be done on
any new problems to see if this remains the case.

\subsubsection{\label{result:upper_bc} Upper BC}

\begin{figure}
\epsscale{.75}
\plotone{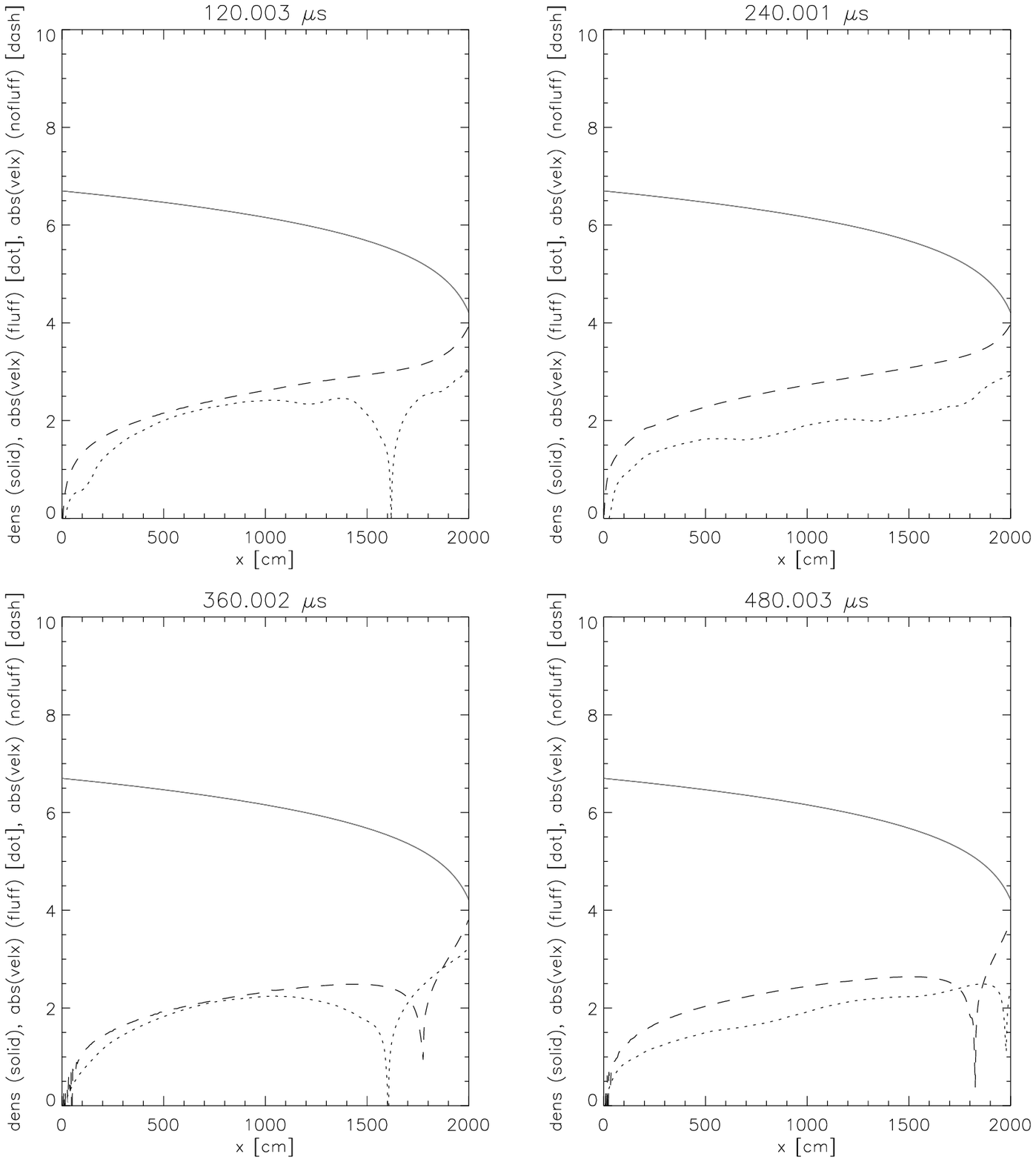}
\caption{\label{fig:iso_fluff_comp} A comparison of the velocities
resulting when the top boundary is hydrostatic (dashed line) verses a
buffer of low density `fluff' (dotted line) for the isothermal
atmosphere initial model.  The four panels show the profiles at 120,
240, 360, and 480 $\micros$.  We see that the `fluff' case yields
lower velocities throughout the atmosphere.}

\end{figure}

\begin{figure}
\epsscale{.75}
\plotone{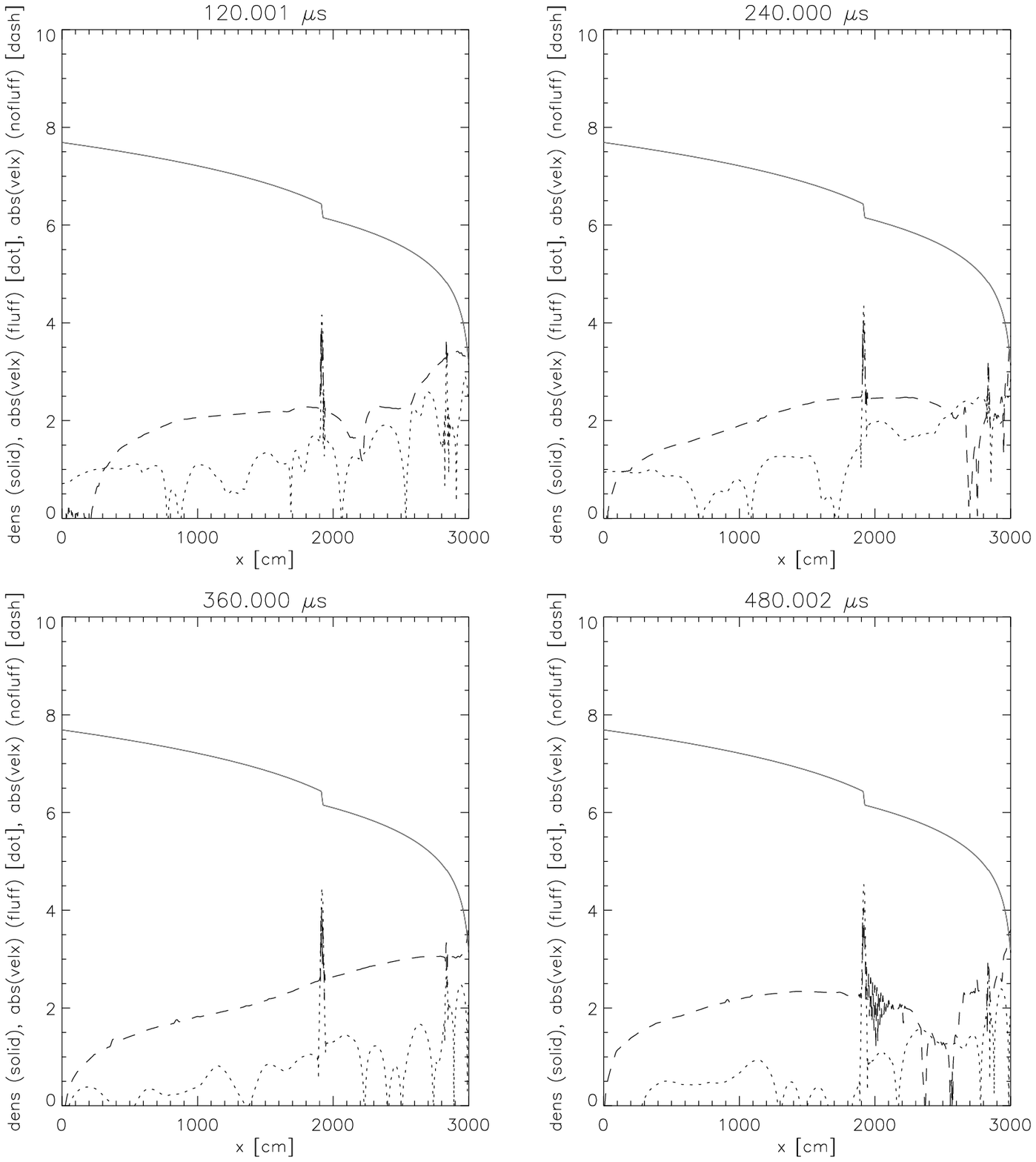}
\caption{\label{fig:xrb_fluff_comp} A comparison of the velocities
resulting when the top boundary is hydrostatic (dashed line) verses a
buffer of low density `fluff' (dotted line) for the Kepler initial
model.  As in the simple isothermal model, the `fluff' boundary
condition yields lower velocities throughout the atmosphere.}

\end{figure}

Figure \ref{fig:iso_fluff_comp} shows the density and velocity as a
function of height at several times for the isothermal atmosphere
initial model, with both the fluff upper boundary and a hydrostatic
boundary at the top of the domain.  The hydrostatic boundary was first
order, constant temperature, with a diode condition on the velocity.
The isothermal, complex EOS initial model was used for these runs.
The two runs were done with the same spatial resolution, on a uniform
mesh.  In the fluff case, the domain extended to 4000 cm, but we only
show the lower 2000 cm, to match the domain used in the hydrostatic
upper boundary case.  Figure \ref{fig:xrb_fluff_comp} shows the same
comparison for the Kepler initial model, again run at identical
spatial resolutions, with domain sizes of 3000 cm (without fluff) and
6000 cm (with fluff).

In both figures we see that the fluff boundary at the top of the
domain is at least as effective as a hydrostatic boundary at the top in
maintaining a hydrostatic atmosphere.  In fact, the velocities
generated throughout the atmosphere with the fluff boundary condition
are generally smaller in magnitude than those with the upper
hydrostatic boundary.  Coupled with the benefit that the envelope is
able to expand on our grid with out losing dynamically important mass
through the top boundary, the fluff condition at the upper boundary is
the optimal choice.

\subsection{Effect of Initialization Method}

\begin{figure}
\epsscale{.95}
\plotone{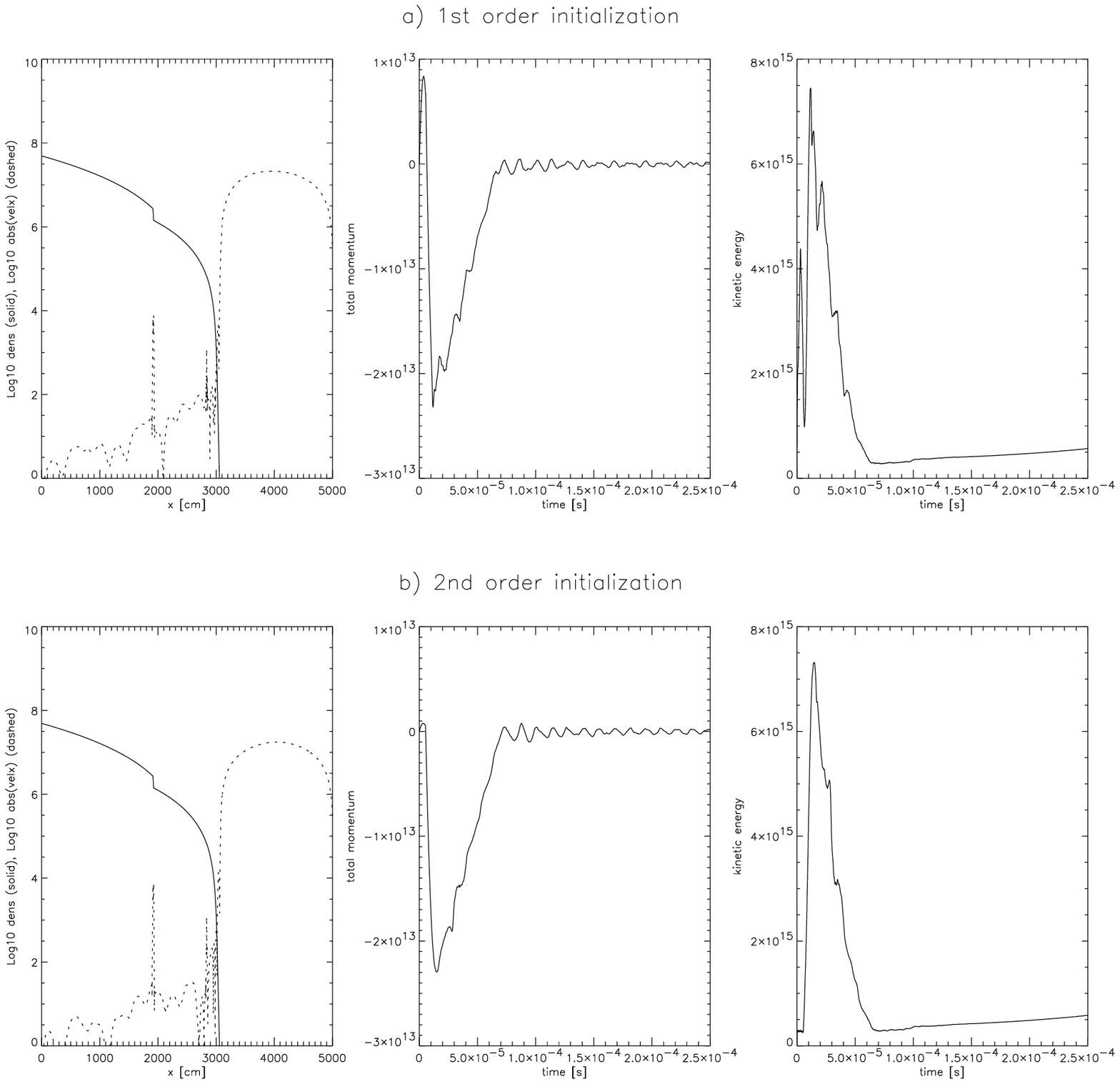}
\caption{\label{fig:init_order_comp} Comparison of the 
order of the HSE differencing used at the initialization stage for the
Kepler initial model.  Both runs used a hydrostatic lower boundary with 
constant temperature and diode velocity condition.   Note that the 
higher-order initialization has a much ($\sim \times 8$) quieter initial
transient, but that medium- and long-term evolution is identical.  This
is as expected unless there is physics (eg., burning) in the simulation 
whereby the initial transients can feed back into the long-term evolution.}

\end{figure}

All of the boundary condition comparisons shown above were done with
second order differencing in the initialization.  We can also use the
first order differencing (Eq. \ref{eq:firstorderdiff}) to put the
initial model into HSE on our grid.  This does not have nearly as
large an effect as the different boundary conditions do.  Figure
\ref{fig:init_order_comp} shows the Kepler initial model with both
first and second order differencing done at initialization.  Both runs
used the hydrostatic lower boundary with constant temperature and diode
velocities, and the standard PPM algorithm.

Looking at the plot of momentum verses time, we notice that at early
times, the first order case generates higher positive velocities
initially than the second order case, as the model relaxes to the
grid.  The two cases quickly settle down to roughly the same state
after about $50\micros$, and by the end of the calculation
($250\micros$), the plot of velocity as a function of height does not
show much difference between the two runs.  This is to be expected;
absent any driving terms from poor boundary conditions or energy
inputs, the simulation will eventually `find' hydrostatic equilibrium
on its own.  However, a good choice of initial model will strongly
reduce the initial transients.  This model is more complicated than
the simple isothermal atmosphere because of the composition interface.
This leads to the more severe velocities.  We note that this
comparison was done with the best choice boundary conditions we had
available.  If the boundary conditions are poor, or the problem is
under resolved, then spurious velocities will continue to be driven in
the simulation every timestep.  Everything else, no matter how bad it
is, will eventually even out as the Euler equations do their work and
find HSE.  The order of the initialization method should match the
order of the boundary condition in to avoid transients caused by jumps
at the boundaries.

In some problems, the initial transient may not be a problem.  In
others, however, they may interfere with a phenomenon one is trying to
measure, or --- if the transients cause spurious burning, or cause
material unphysically to leave your computational domain --- they may
alter or change entirely the long-term evolution of the system.

\subsection{\label{res_effect} Effect of Resolution}

\begin{figure}

\plotone{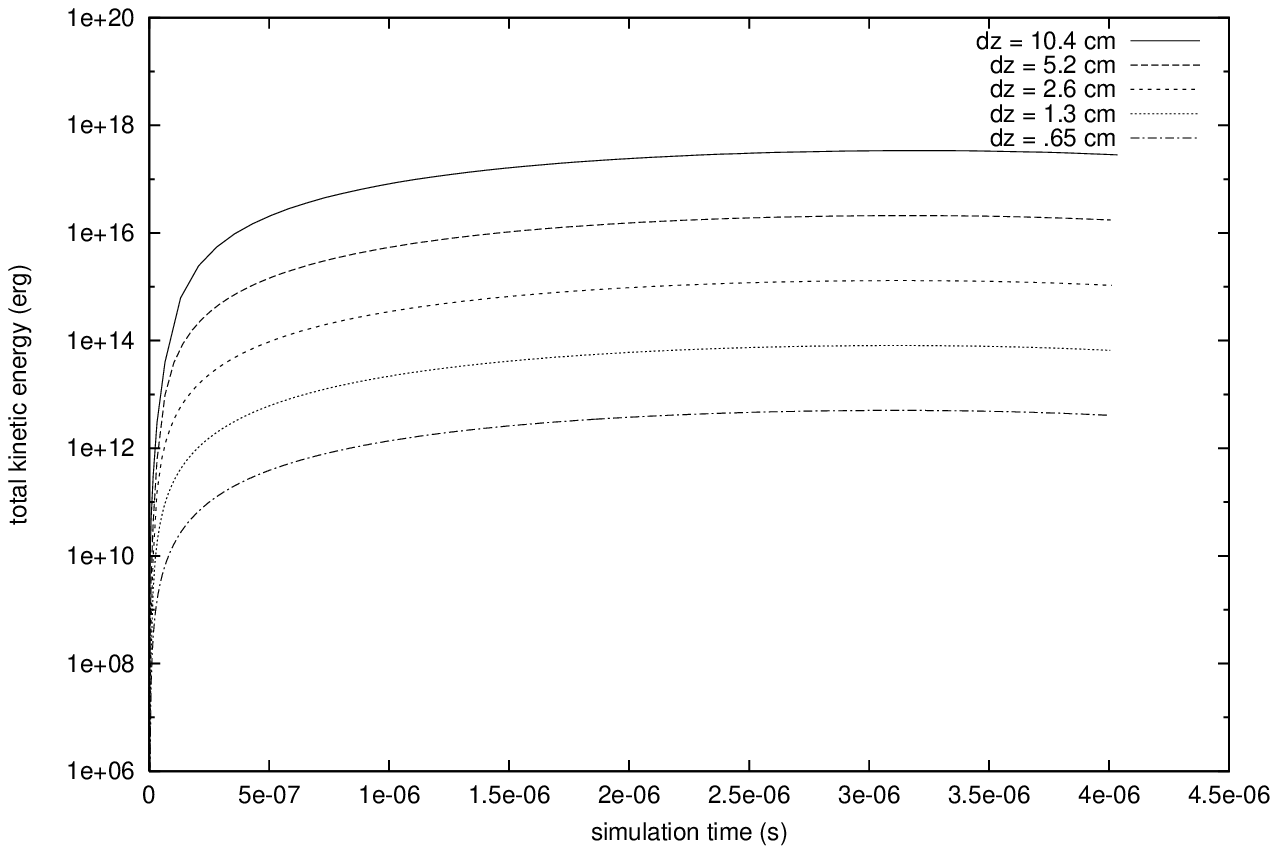}
\caption{\label{convergence} \label{fig:ppm-nohse} 
         Plots of kinetic energy for the isothermal
         atmosphere described in \S \ref{realeos_atm}.
         Since this model should be hydrostatic, the `correct' KE is zero.
         The second-order differencing scheme was used for initialization,
         and second-order constant-temperature boundary conditions were
         used with diode velocity conditions.  For scale, the mass in the 
         simulation domain is $\sim 3 \times 10^9~{\mathrm g}$, so that a 
         KE of $10^{15}~{\mathrm{erg}}$ corresponds to a density-weighted 
         RMS velocity of $6 \times 10^2~{\mathrm{cm/s}}$, while the sound 
         speed of the medium is from $1-4\times 10^{8}~{\mathrm{cm/s}}$.
	 }
\end{figure}

\begin{figure}
\plotone{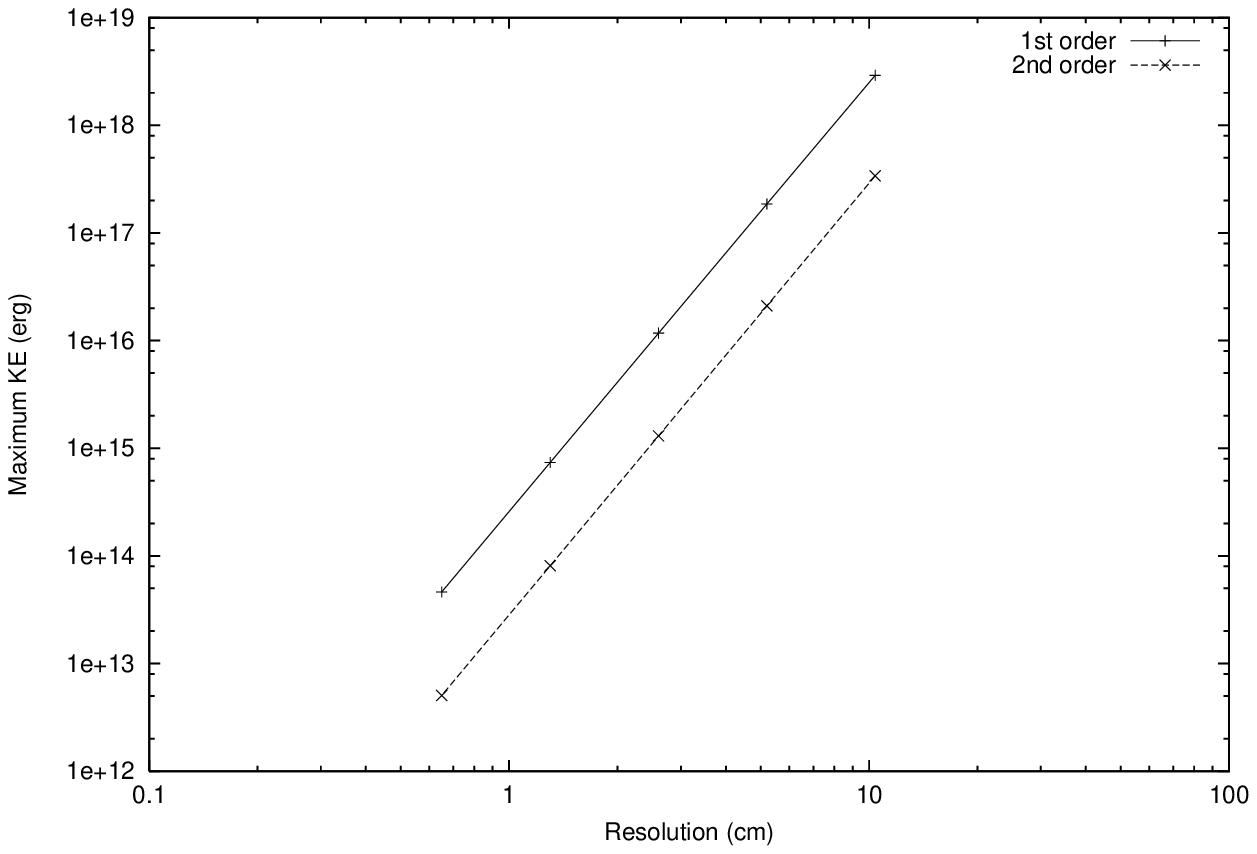}
\caption{\label{convergence1st2ndPPM} The maximum kinetic energy in
          Figure \ref{convergence} (dashed), and with the corresponding
          simulations run with first order BCs and initial differencing
          (solid), both with PPM.   We see in both cases the kinetic energy
          converging as the $4^{\mathrm{th}}$ power of resolution, meaning
          that the velocities are converging as the $2^{\mathrm{nd}}$ power.}
\end{figure}

Figure \ref{convergence} shows the kinetic energy verses time for an
isothermal atmosphere with 2nd-order differencing and 2nd-order
constant-temperature boundary conditions.  This figure shows only the
short term evolution.  Our estimate in \S \ref{gamma_model} showed the
importance of resolving the scale height in minimizing any spurious
velocities.  We therefore expect to see improvement in the ambient
velocities as we increase the spatial resolution.  As we might expect,
our model shows second-order convergence (see Figure
\ref{convergence1st2ndPPM}) in resolution.

We can combine the effects of higher-order initialization and 
resolution by looking at the convergence behavior of the velocities
in the isothermal atmosphere with different initialization schemes.
This is also shown in Figure \ref{convergence1st2ndPPM}.   We see
that with the operator-split gravity, we still only get 2nd order
convergence even with the higher-order spatial discretization for
the initial model, however the spurious kinetic energy is an order
of magnitude lower.
 
Figure \ref{fig:xrb_res_study} shows the long term evolution of the
Kepler initial model for four different spatial resolutions.  The `6
level' curve corresponds to that shown in Figure \ref{fig:bc_plots}.
All four simulations were run with the first order initialization and
the constant temperature hydrostatic boundary conditions.  We see that
as the resolution is increased (each level is a doubling in
resolution), the momentum and kinetic energy decreases dramatically,
as we would expect.

\subsection{Effect of Input States to Riemann Problem}

\begin{figure}
\plotone{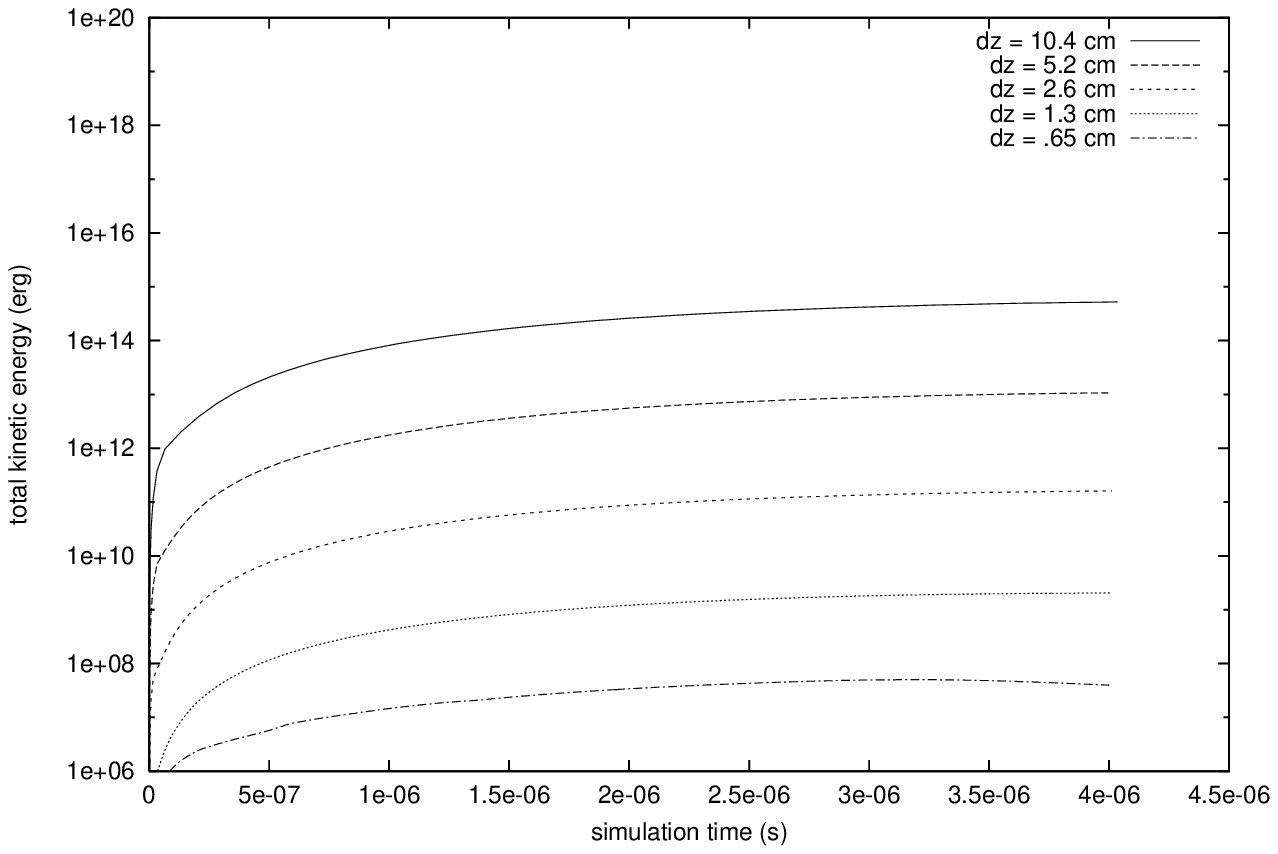}
\caption{\label{fig:ppm-hse} Same as Figure \ref{fig:ppm-nohse}, but with 
          the simulations run with the modified PPM states described in 
          \S \ref{sec:ppmhse}.}
\end{figure}

\begin{figure}
\plotone{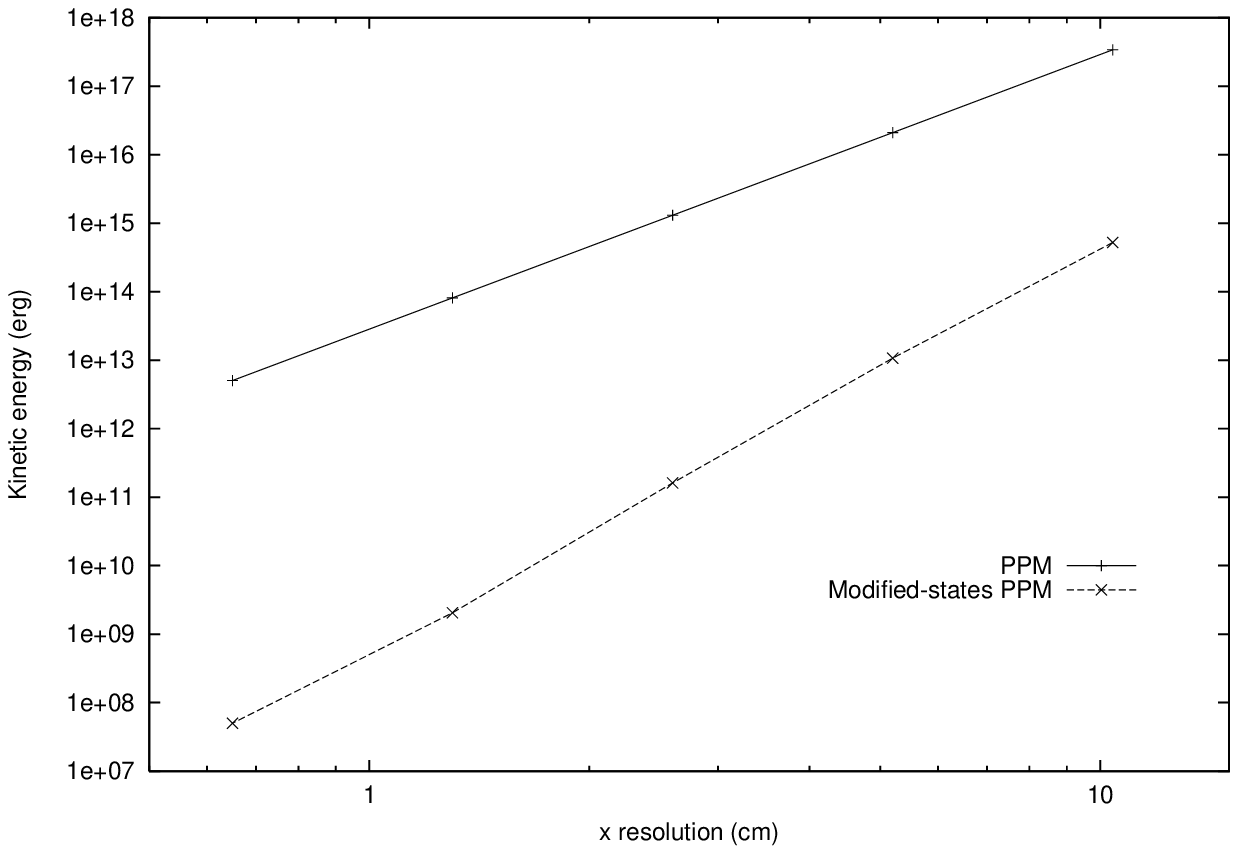}
\caption{\label{fig:ppmhseppmconverge} Same as Figure 
          \ref{convergence1st2ndPPM}, 
          but with the simulations run with the modified PPM states 
          described in \S \ref{sec:ppmhse} added.   We see the second-order
          convergence turning into third-order in the case of the 
          modified-states solver.}
\end{figure}

We performed, with our modified-states PPM, the same numerical
experiment as in \S \ref{res_effect}, which already had fairly low
velocities.  The results are shown in Figure \ref{fig:ppm-hse}.
By modifying the input states, the numerically-induced kinetic energy
drops by three orders of magnitude over those shown in Figure
\ref{fig:ppm-nohse}, and converges to third-order, rather than to
second-order, as shown in Figure \ref{fig:ppmhseppmconverge}.
The more direct coupling of the gravity into the hydrodynamical
solver increases both the accuracy of the solution and the convergence
properties.

This improvement will only become apparent when using a good
initialization method and boundary condition, otherwise, any
improvement it may yield will be swamped by noise from these other sources.
We used the second order constant temperature/diode velocity
boundaries for these simulations to minimize any boundary effects.

\section{Conclusions}

We have studied the effects of the initialization process, boundary
conditions, resolution, and the hydrodynamic algorithm on the process
of mapping an initial model onto a Eulerian grid in a Godunov-type code.
We saw that depending on how one goes about the process, the result can
be a nicely relaxed, quiet mapping, or one dominated by high velocities,
swamping out any physical processes that one may be interested in
studying.

Boundary conditions have the greatest impact on maintaining the
stability of the hydrostatic atmosphere.  The standard reflecting
boundary is poorly suited for maintaining a hydrostatic envelope---the
inability of minor pressure disturbances to escape off the bottom of
the grid as the model relaxes results in high velocities.  The actual
choice of boundary condition should reflect the physics of the
atmosphere being supported.  If the initial model is isothermal, then
an isothermal/HSE boundary condition provides with best match.  The
constant entropy boundary may be a better match for an atmosphere that
is generated by a stellar evolution code.  For any model, a
hydrostatic boundary condition, using differencing that matches that
of the initialization, with a secondary constraint that matches the
physics of the model atmosphere is the best solution.  Assuming that the
density is constant in the boundary proved to be the worst assumption.
The degeneracy of the EOS puts strong demands on the temperature
profile to counteract the weight of the atmosphere.  This assumption
may fare better with a gamma-law EOS, but this was not considered in
the present paper.

We also demonstrated that, once an appropriate boundary condition is
chosen, the treatment of source terms in the hydrodynamic solver can
have an impact on the stability of the atmosphere.  Our
modified-states PPM was effective in reducing the magnitude of the
spurious velocities generated as our model relaxed onto our grid.
This method, or that proposed by \citet{leveque}, can be adopted to
most Godunov type codes to increase the accuracy of the hydrodynamics
in the presence of gravity.

As expected, the resolution is important in reducing the errors in
maintaining a hydrostatic atmosphere.  The resolution should be chosen
to be fine enough to resolve the scale height of the atmosphere well,
and to keep ambient velocities smaller in magnitude than whatever
physical processes under study would yield.   In an AMR calculation,
the coarsest resolution used in the simulation must be this fine, even
in regions of the atmosphere with no features.

Boundary conditions, source terms, and poor resolution can all be
sources of spurious velocities throughout a calculation.   By contrast,
initialization methods can cause at most a transient while the
simulation settles into hydrostatic equilibrium.   This transient may
take a while to settle, however, or may have dynamical consequences
later in the simulation, so care must be taken.  The initialization
methods described in this paper produce profiles which generate very
small transients.

\begin{figure}
\epsscale{.6}

\includegraphics[angle=90,scale=.65]{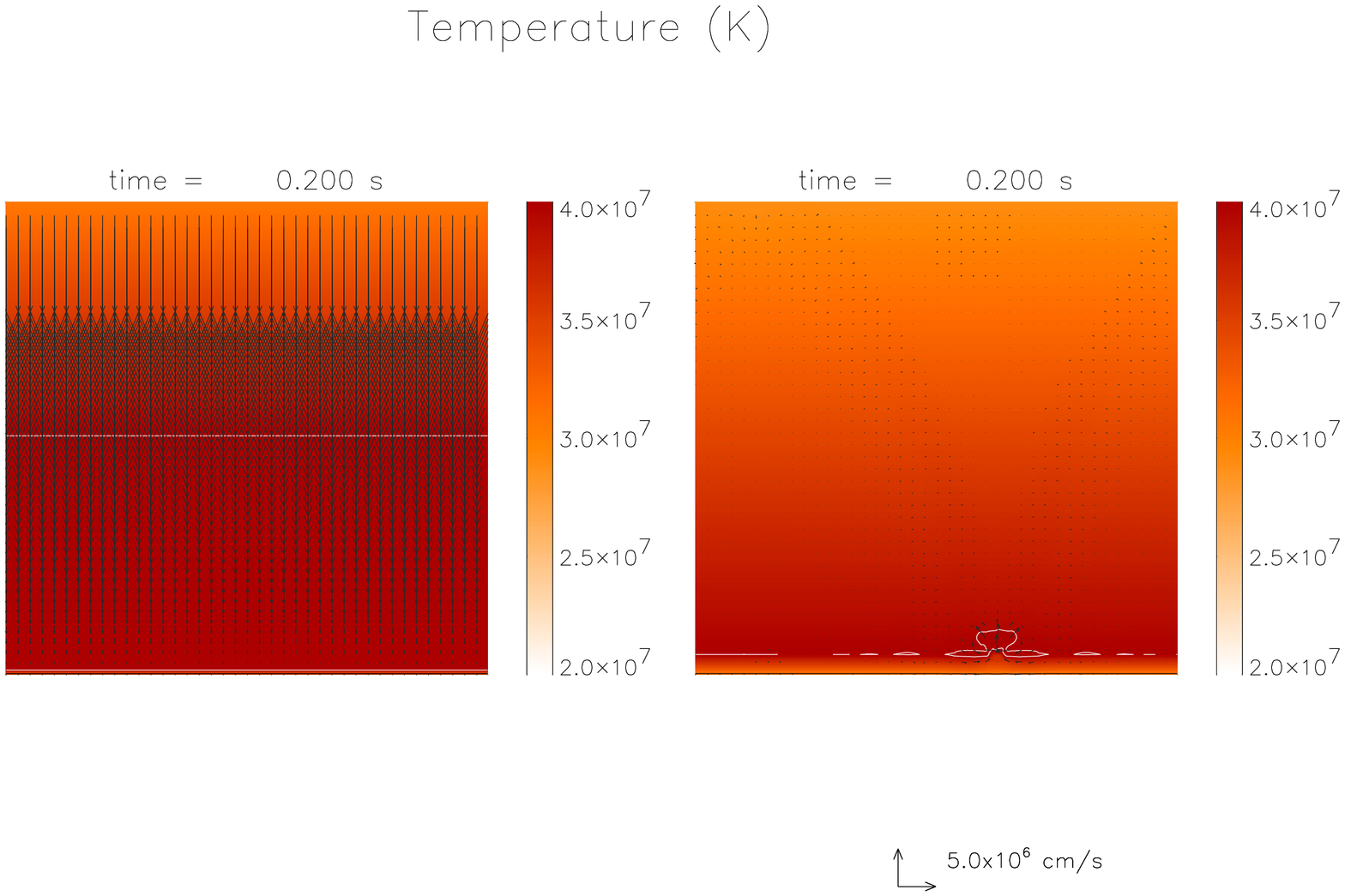}
\caption{\label{fig:beforeafter}  Two-dimensional simulations of a perturbed
               nova precursor atmosphere, zoomed in to near the interface, 
               evolved 0.20 s.   Plotted is temperature with velocity
               vectors.  A white contour shows the highest
               temperature region ($T > 4.01 \times 10^7 {\mathrm{K}}$), and
               a black contour shows the interface between the C/O white dwarf
               and the accreted material.
	       Above are the simulation results without adjusting the
	       profile to HSE in this code, with reflecting boundary
	       conditions, and without the modified-states PPM.  Below
	       are results with the second-order differencing to bring
	       the model to HSE, second-order constant-temperature HSE
	       boundary conditions, and using the modified-states PPM.
	       Note the difference in velocity magnitudes, spurious
	       compression of the bottom layer, and resulting
	       compressional heating, which wipes out the real physical
               effect under consideration, the initiation of convection.}
\end{figure}

We note that the methods that we describe in this paper can be applied
to any initial hydrostatic atmosphere.  As a final example, we show the
havoc a poorly initialized model can wreak in a simulation.  Figure
\ref{fig:beforeafter} represents two-dimensional simulations evolved
from a one dimensional model provided by S.~A.~Glasner (2002, private
communication) of a classical nova precursor; it is a model from the same
simulation that produced the 1d model used in \cite{glasner,kercek},
but at an earlier time, before convection begins.  Thus, one could
hope to examine the multidimensional onset of convection in these
nova precursors \citep{novapaper}.  The convection is driven by nuclear
reactions `simmering' near the interface between the C/O white dwarf
and the accreted layer of stellar material.  Since the vertical scale
of the simulation is significant compared to the radius of the white dwarf,
plane-parallel inverse-square gravity is used rather than constant gravity.

The model is perturbed in the highest-temperature region of the accreted
atmosphere with a 10\% temperature increase at time $t = 0$.  The white
contours in the figures enclose the region in the simulation with a
temperature greater than this perturbed temperature.   The black contour
marks the white dwarf/accreted material interface.

Without taking any of the precautions outlined in this paper, an
unphysical `settling' occurs, caused by poor boundary conditions and
differences in the EOS.  This generates large velocities ($v \sim
5\times 10^6 {\ \mathrm{cm\,s^{-1}}}$) and compressional heating, as
shown in the figure.  The heating, combined with unphysical mixing
across the interface caused by the large velocities, then cause a
completely spurious layer of increased burning, which then dominates
the long-term evolution of the simulation.    By contrast, the
simulation using constant-temperature second-order boundary conditions,
second-order initialization, and the modified-states PPM shows the
beginning of the formation of convective rolls.  Both simulations were
done on a $1920 \times 640$ uniform mesh, with a computational domain
of $1 \times 10^7 {\ \mathrm{cm}} \times 3 \times 10^7
{\ \mathrm{cm}}$; Figure \ref{fig:beforeafter} shows only the domain
near the interface of the white dwarf and the accreted material.

\acknowledgements 

The authors thank Stan Woosley for providing initial models from the
Kepler code for use in this study, and Ami Glasner for providing the
initial nova model used.  We thank Edward Brown for useful
comments on the manuscript.  Support for this work was provided by the
Scientific Discovery through Advanced Computing (SciDAC) program of
the DOE, grant number DE-FC02-01ER41176 and DOE grant number B341495 to
the ASCI/Center for Astrophysical Thermonuclear Flashes at the
University of Chicago.  LJD is supported by the Krell Institute CSGF.
The work of T. Plewa was partly supported by the grant 2.P03D.014.19
from the Polish Committee for Scientific Research.
These calculations were performed on the Nirvana Origin 2000 Cluster
at Los Alamos National Laboratory.  All simulations were performed
with the FLASH 2.1.  FLASH is freely available at
http://flash.uchicago.edu/.

\clearpage

\section*{Appendix: Non-Constant Gravity}
In this paper, we have considered only the case of constant gravity.   
It is relatively straightforward to extend this to the case of fixed
but non-spatially-constant gravitational acceleration.

Consider for instance Eq. (\ref{eq:firstorderdiff}).  This was derived
by using a linear reconstruction for the $\avgsub{\rho}{i}$ data and
putting into the equation of hydrostatic balance (Eq. \ref{eq:hse}).
If instead, the quantity $(\rho g)$ is reconstructed, and the same
procedure applied, one ends with
\begin{equation}
\avgsub{P}{+1}-\avgsub{P}{0}  = \frac{\delta}{2} \left ( \avgsub{\rho g}{+1} + \avgsub{\rho g}{0} \right ) \enskip .
\end{equation}
However, to iterate on the equation of state and to calculate a local
average density, one now must deconvolve $\avgsub{\rho g}{1}$ to find
$\avgsub{\rho}{1}$.  There are a number of ways one could do this.  One
way is to use a cell-averaging algorithm consistent with PPM:
\begin{equation}
\label{eq:simpsonsrule}
\avg{f} = \frac{1}{6} \left [ f_L + 4 f_C + f_R \right ] \enskip ,
\end{equation}
where the subscripts $(L,C,R)$ refer to the values of the reconstructed
function at the left cell interface, cell center, and right cell interface,
respectively.   Presumably, whatever our fixed gravitational acceleration,
we can evaluate it numerically at any given point, so that we can then use
\begin{equation}
\label{eq:simpsonsrulegrav}
\avgsub{\rho g}{1} = \frac{1}{6} \left [ \rho\left(\frac{\delta}{2}\right) g\left(\frac{\delta}{2}\right) + 4 \rho\left(\delta\right) g\left(\delta\right) + \rho\left(\frac{3 \delta}{2}\right) g\left(\frac{3 \delta}{2}\right) \right ] \enskip ,
\end{equation}
with the computed values of $g$ and evaluate the pointwise values of
$\rho$ using the (here, linear) reconstruction of density to solve for
the local density average:
\begin{equation}
\avgsub{\rho}{1} = 
     {\frac{\avgsub{\rho}{0}\left(g_R - g_L\right) + 
         12\avgsub{\rho g}{1}}{8g_C + g_L + 3g_R}} \enskip .
\end{equation}

One can repeat this procedure from Eq. (\ref{eq:simpsonsrulegrav}) using
the quadratic reconstruction to get
\begin{equation}
\avgsub{\rho}{1} = 
     {\frac{{\avgsub{\rho}{0}}\left(7 g_R - 2 g_C - 5 g_L\right) + \avgsub{\rho}{-1}\left(-2 g_R + g_C + g_L\right) + 36{\avgsub{\rho g}{1}} }{23{g_C} + 
         2{g_L} + 11{g_R}}} \enskip .
\end{equation}

Thus, one can use the first- or second-order differencing in both the
initialization and the boundary condition to find $\avgsub{\rho g}{1}$,
and then the above equations to find the implied value of $\avgsub{\rho}{1}$,
and iterate on the equation of state as before.

\clearpage

\begin{table}
\caption{\label{parameters} Summary of the different parameters used in the grid of calculations}

\begin{tabular}{ll}

\tableline
\tableline
property & value \\
\tableline
{\em initial model}         & isothermal \\
                            & kepler model \\
\tableline
{\em initialization method} & first order differencing \\
                            & second order differencing \\
\tableline
{\em hydrodynamics}         & ppm \\
                            & ppm/modified-states \\
\tableline
{\em levels of refinement}  & 5 \\
                            & 6 \\
\tableline
{\em boundary conditions}   & reflect \\
                            & constant $\rho$, first order \\
                            & constant $T$, first order \\
                            & constant $s$, first order \\
                            & constant $\rho$, second order \\
                            & constant $T$, second order \\
                            & constant $s$, second order \\
\tableline

\end{tabular}

\end{table}

\clearpage

\clearpage

\clearpage

\clearpage

\clearpage

\clearpage

\clearpage

\clearpage

\clearpage

\clearpage

\begin{figure}

\epsscale{0.95}
\plotone{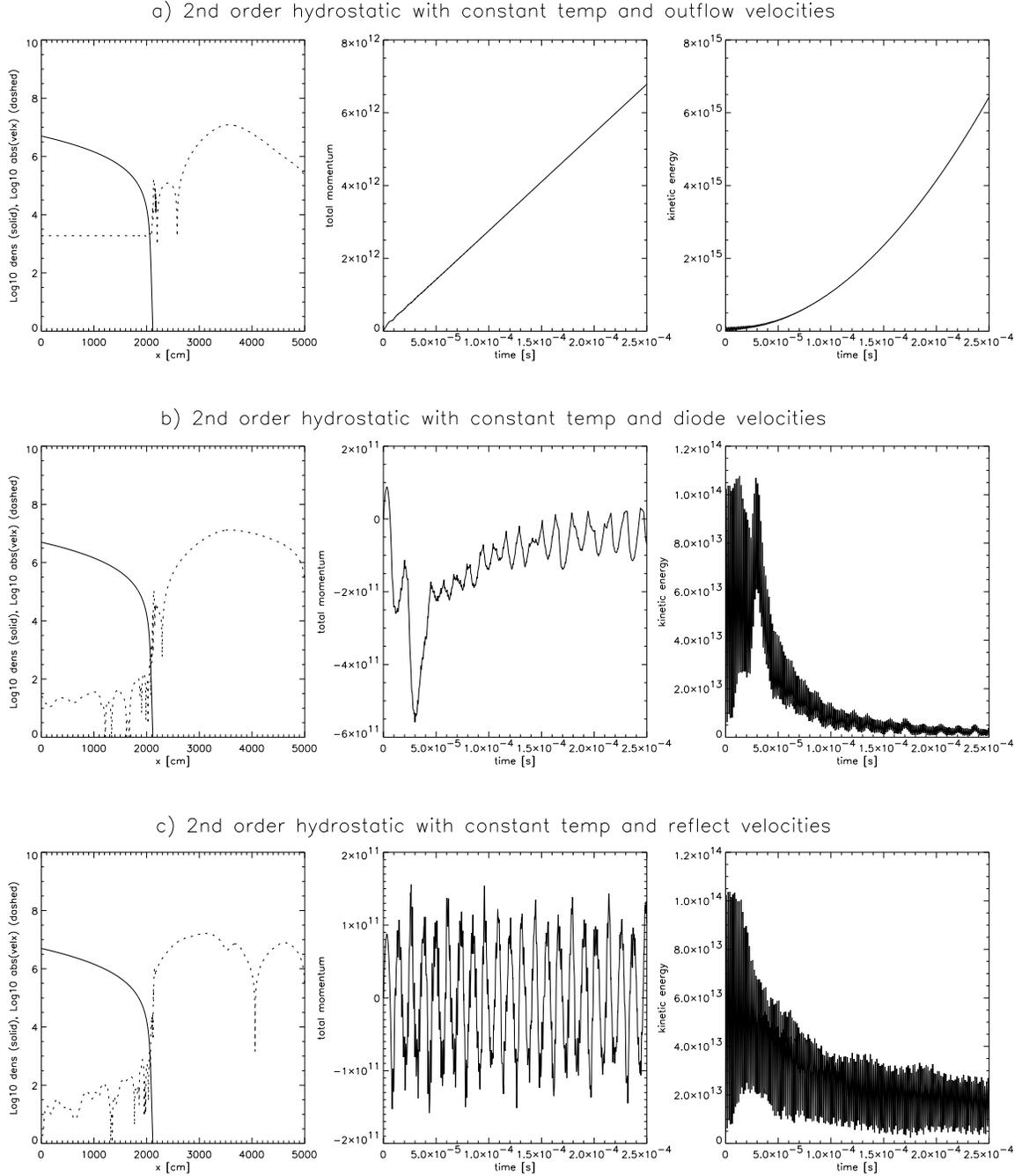}
\caption{\label{fig:iso_bc_plots} Effect of the boundary conditions on
the velocity for the isothermal initial model.  Each simulation was
run for $250 \micros$ with a spatial resolution of $2.4 \
\mathrm{cm}$.  The left panel shows the density and velocity as a
function of height at $250 \micros$.  The middle panel shows the total
momentum as a function of time, and the right panel shows the kinetic
energy as a function of time.}

\end{figure}

\begin{figure}
\figurenum{\ref{fig:iso_bc_plots}}
\epsscale{0.95}
\plotone{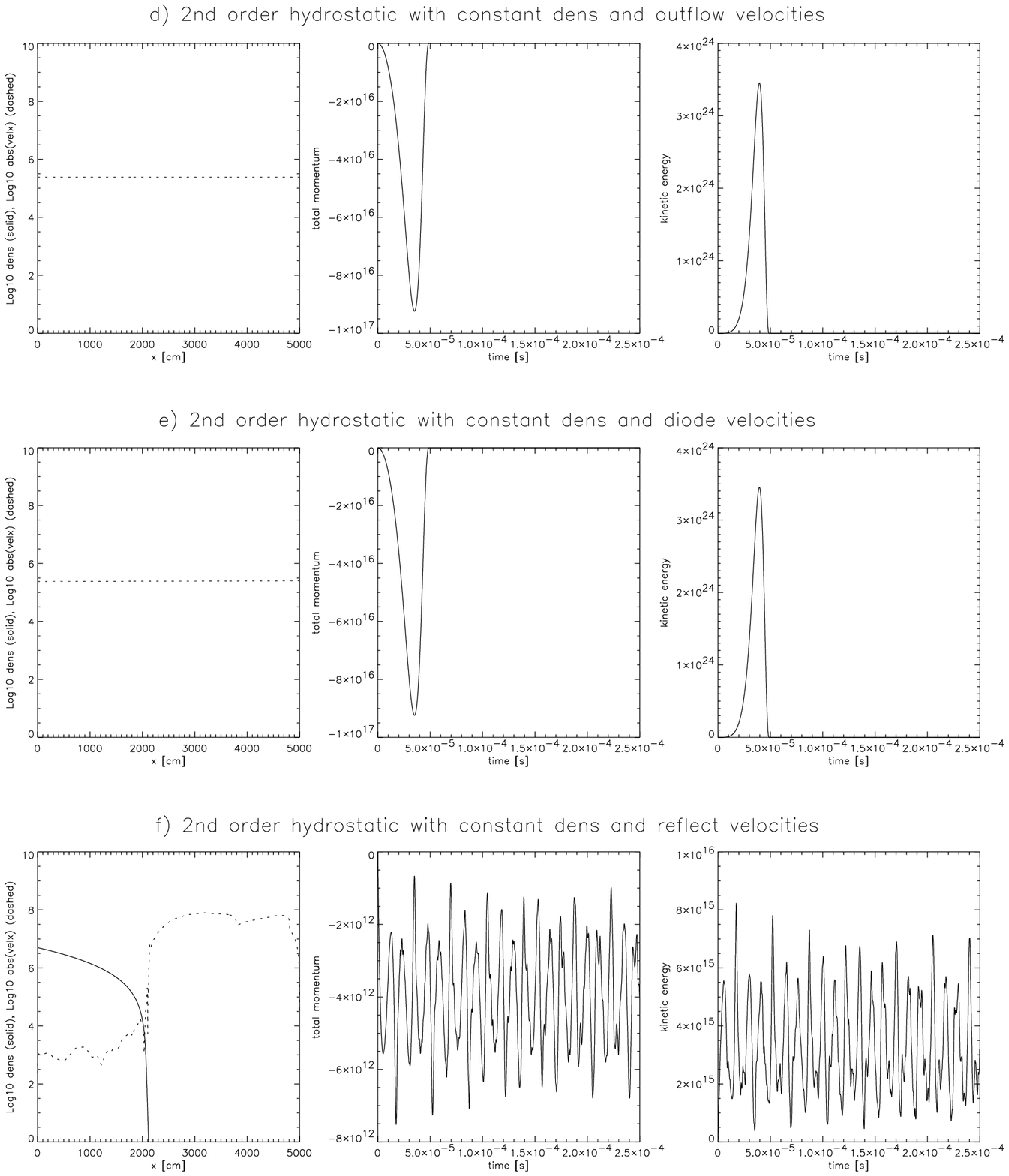}
\caption{cont.}

\end{figure}

\clearpage

\begin{figure}
\figurenum{\ref{fig:iso_bc_plots}}
\epsscale{0.95}
\plotone{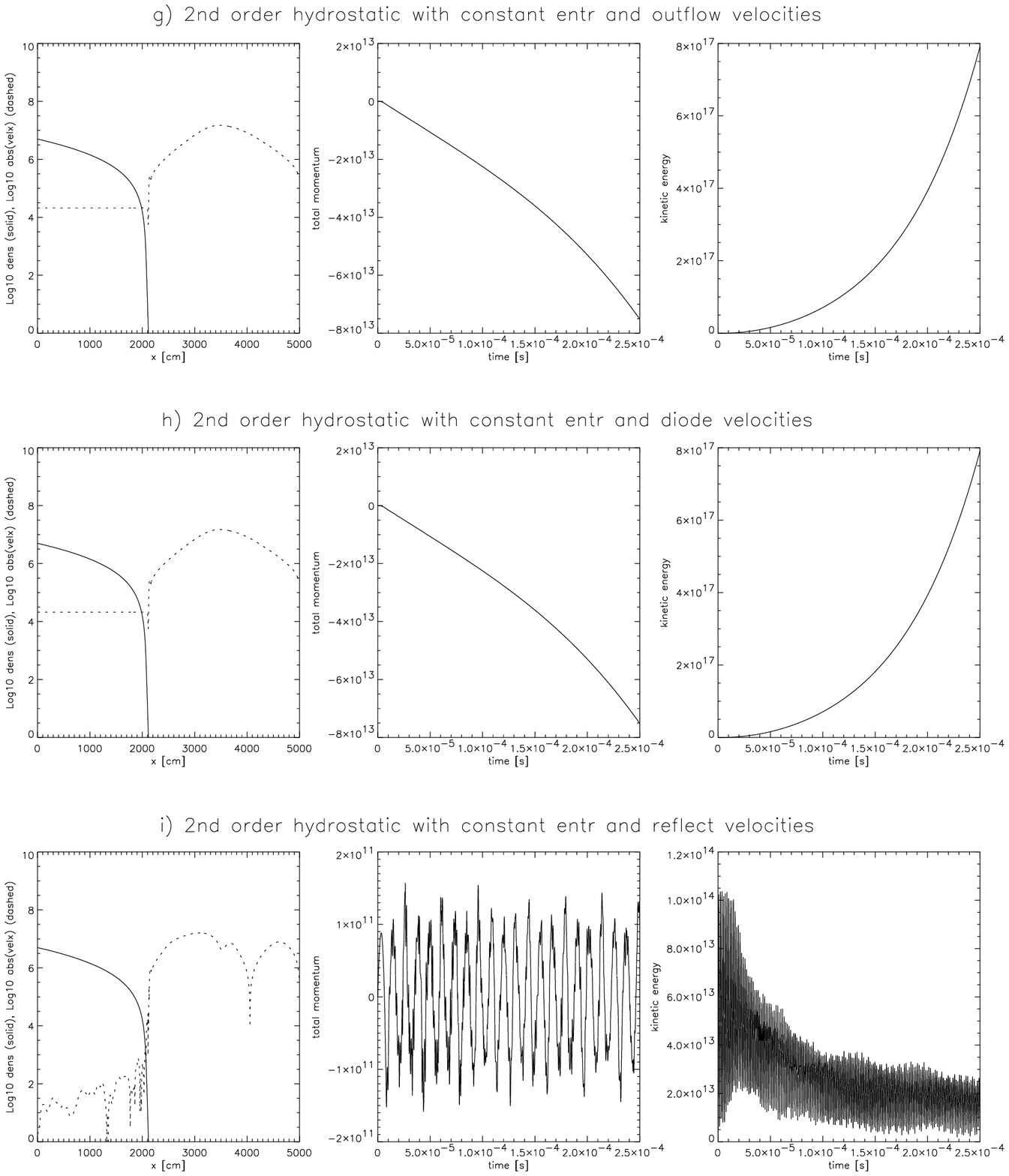}
\caption{cont.}

\end{figure}

\clearpage

\begin{figure}
\figurenum{\ref{fig:iso_bc_plots}}
\epsscale{0.95}
\plotone{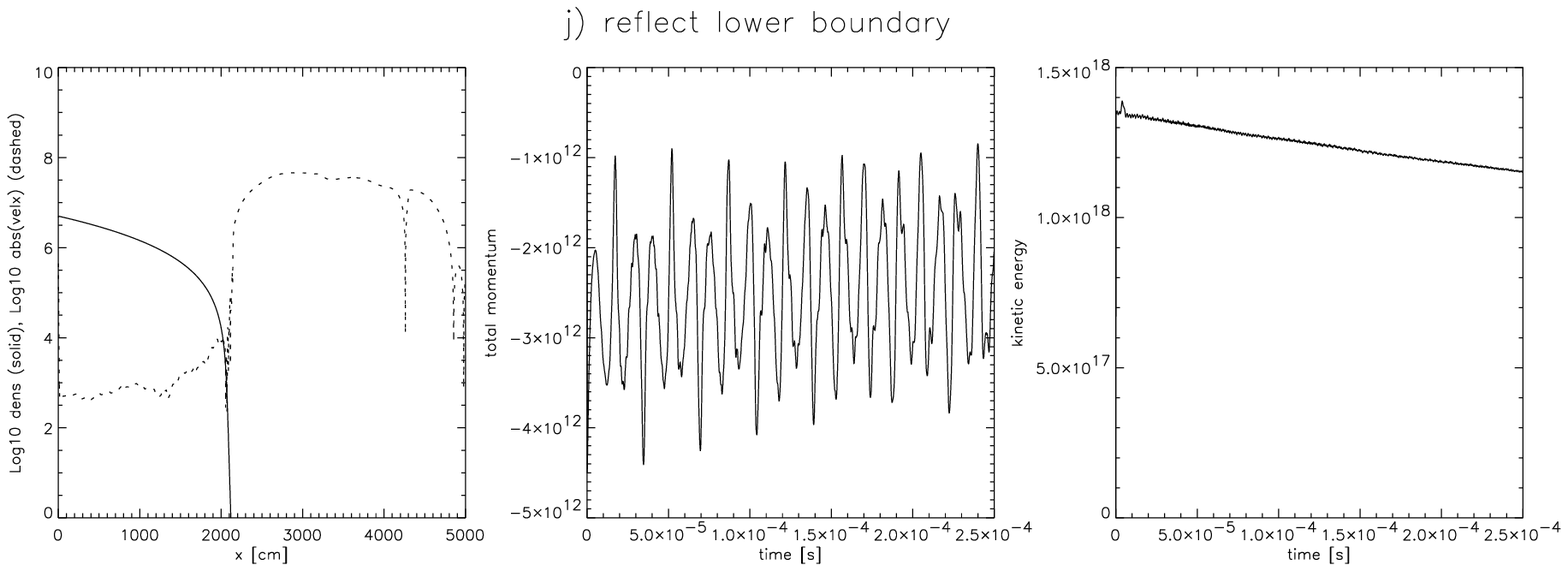}
\caption{cont.}

\end{figure}

\clearpage

\begin{figure}

\epsscale{0.95}
\plotone{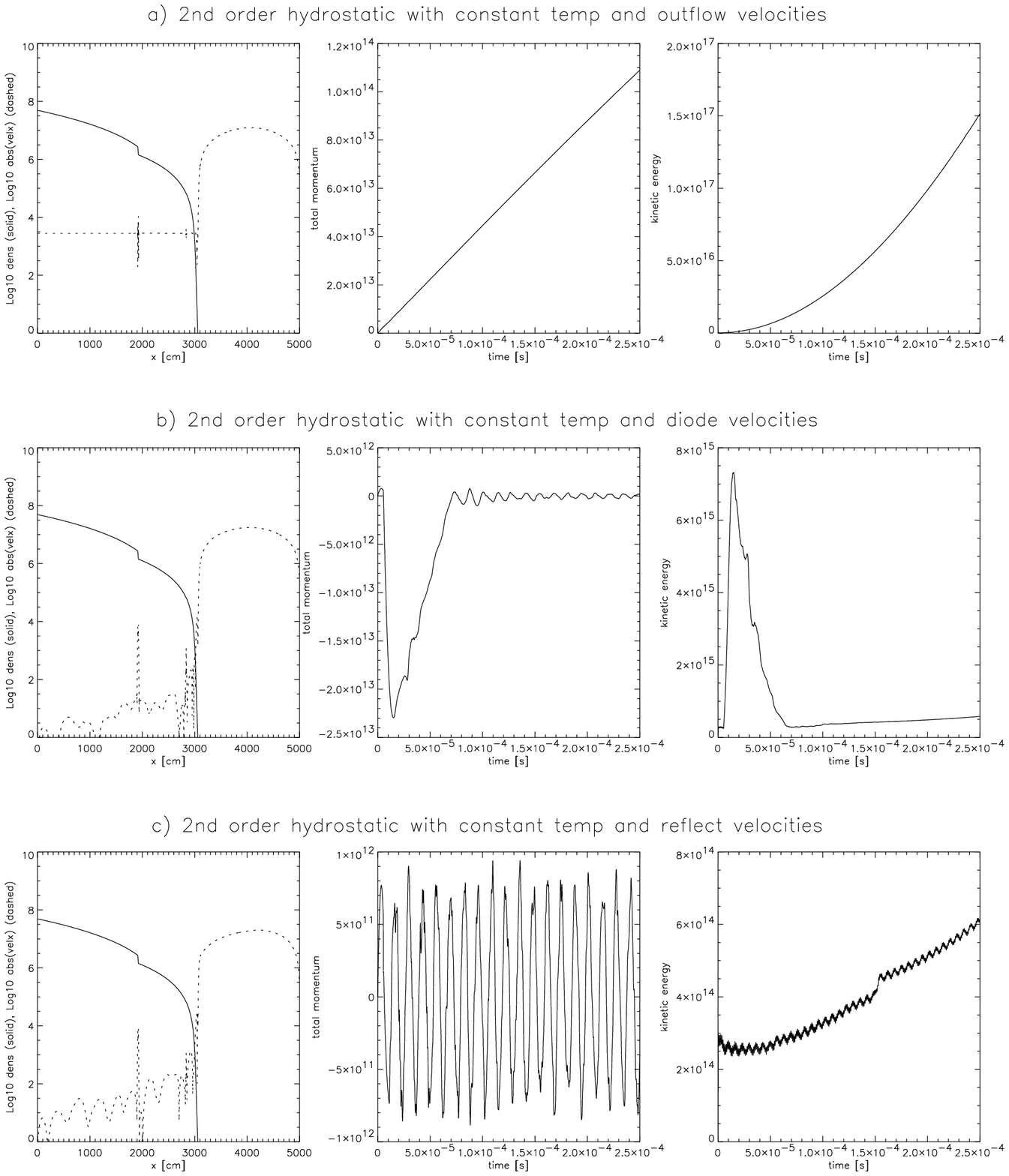}
\caption{\label{fig:bc_plots} Effect of the boundary conditions on the
velocity for the Kepler initial model.  For each choice of boundary
condition, three plots are shown: a snapshot in time of the velocity
structure vs.\ height (taken at $250 \micros$); a plot of the total
momentum vs.\ time, and a plot of the total kinetic energy vs.\ time.}

\end{figure}

\begin{figure}
\figurenum{\ref{fig:bc_plots}}
\epsscale{0.95}
\plotone{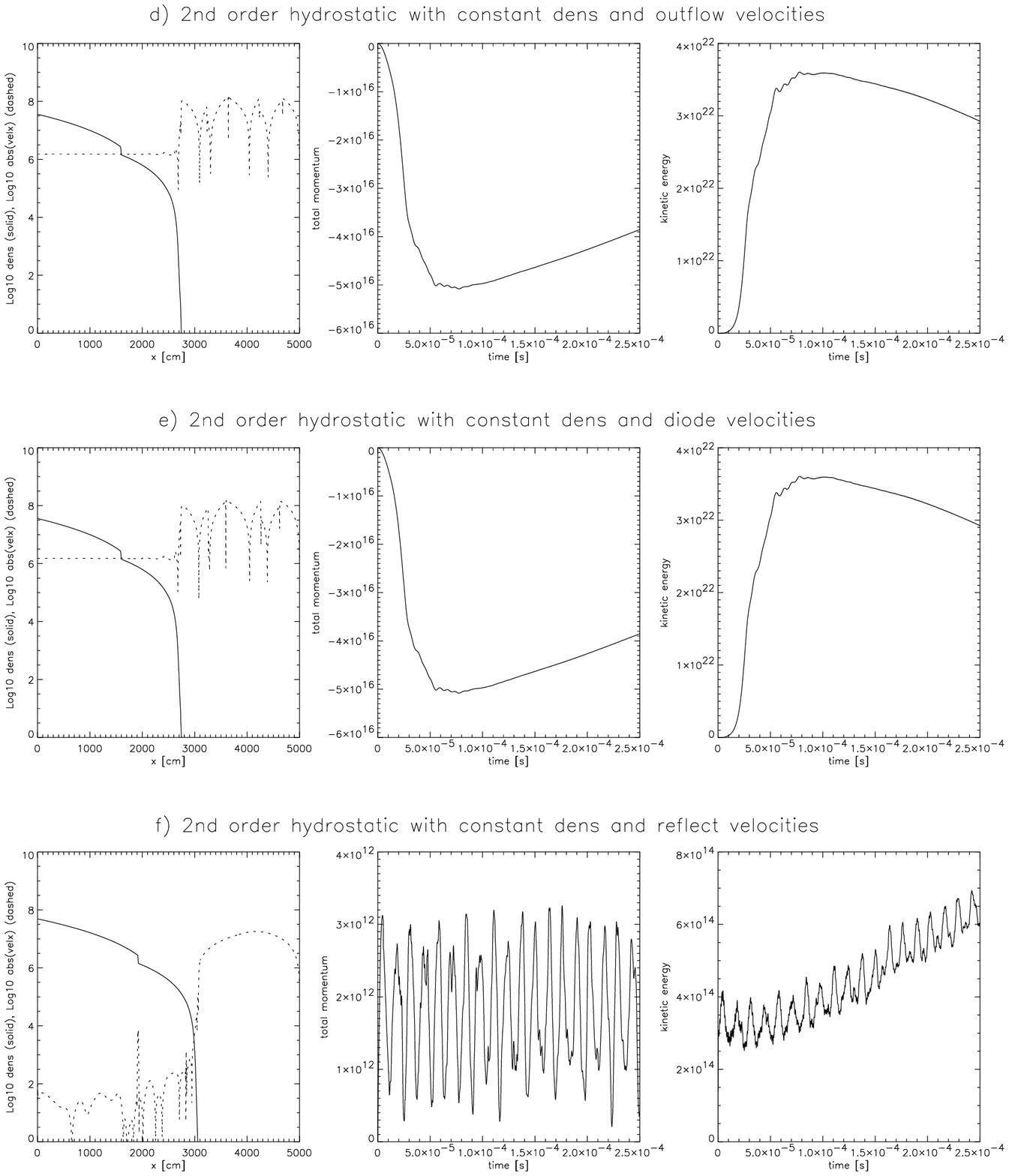}
\caption{cont.}
\end{figure}

\begin{figure}
\figurenum{\ref{fig:bc_plots}}
\epsscale{0.95}
\plotone{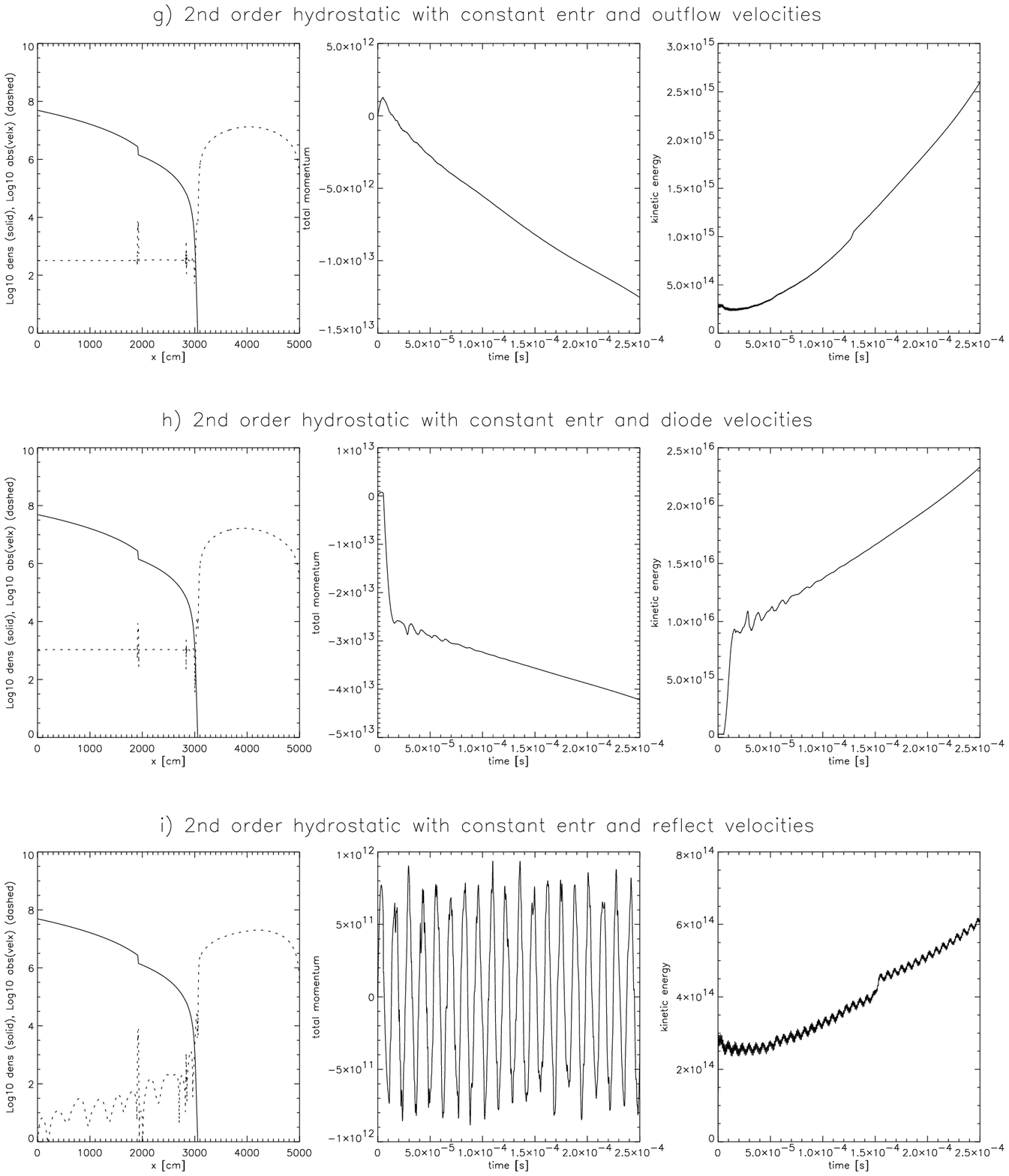}
\caption{cont.}
\end{figure}

\begin{figure}
\figurenum{\ref{fig:bc_plots}}
\epsscale{0.95}
\plotone{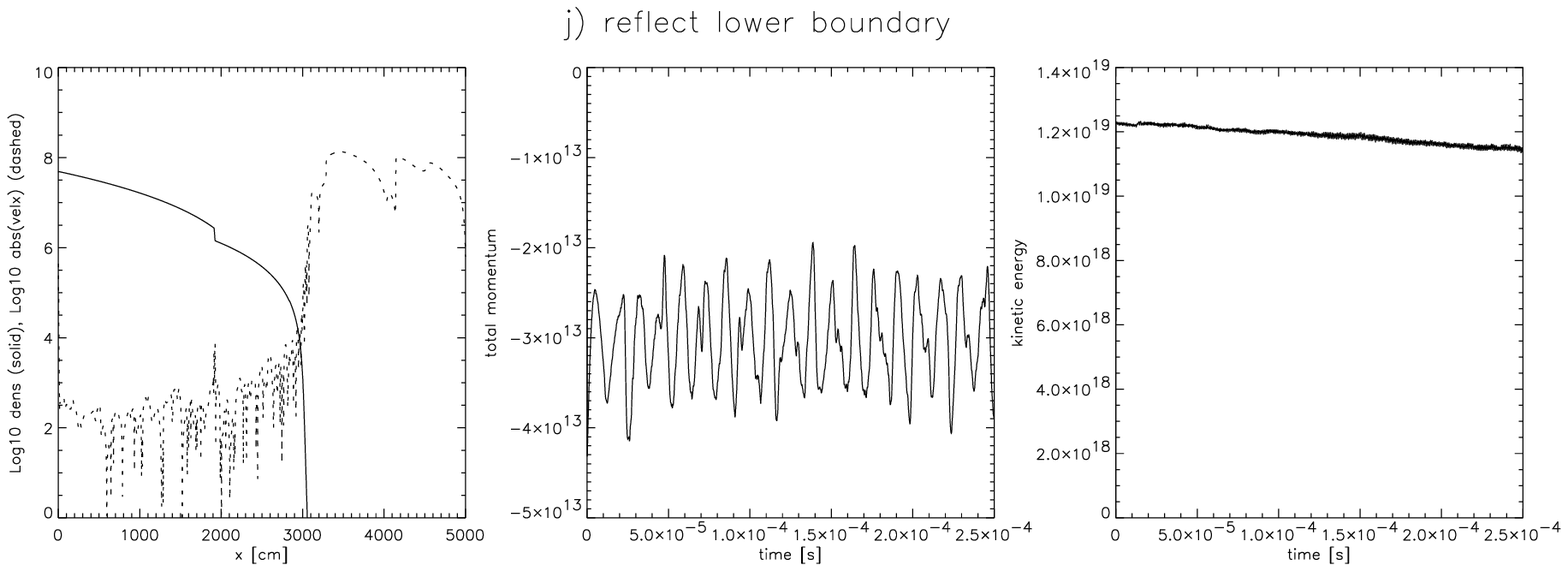}
\caption{cont.}
\end{figure}

\clearpage

\clearpage

\clearpage

\clearpage

\clearpage

\clearpage

\begin{figure}

\epsscale{0.95}
\plotone{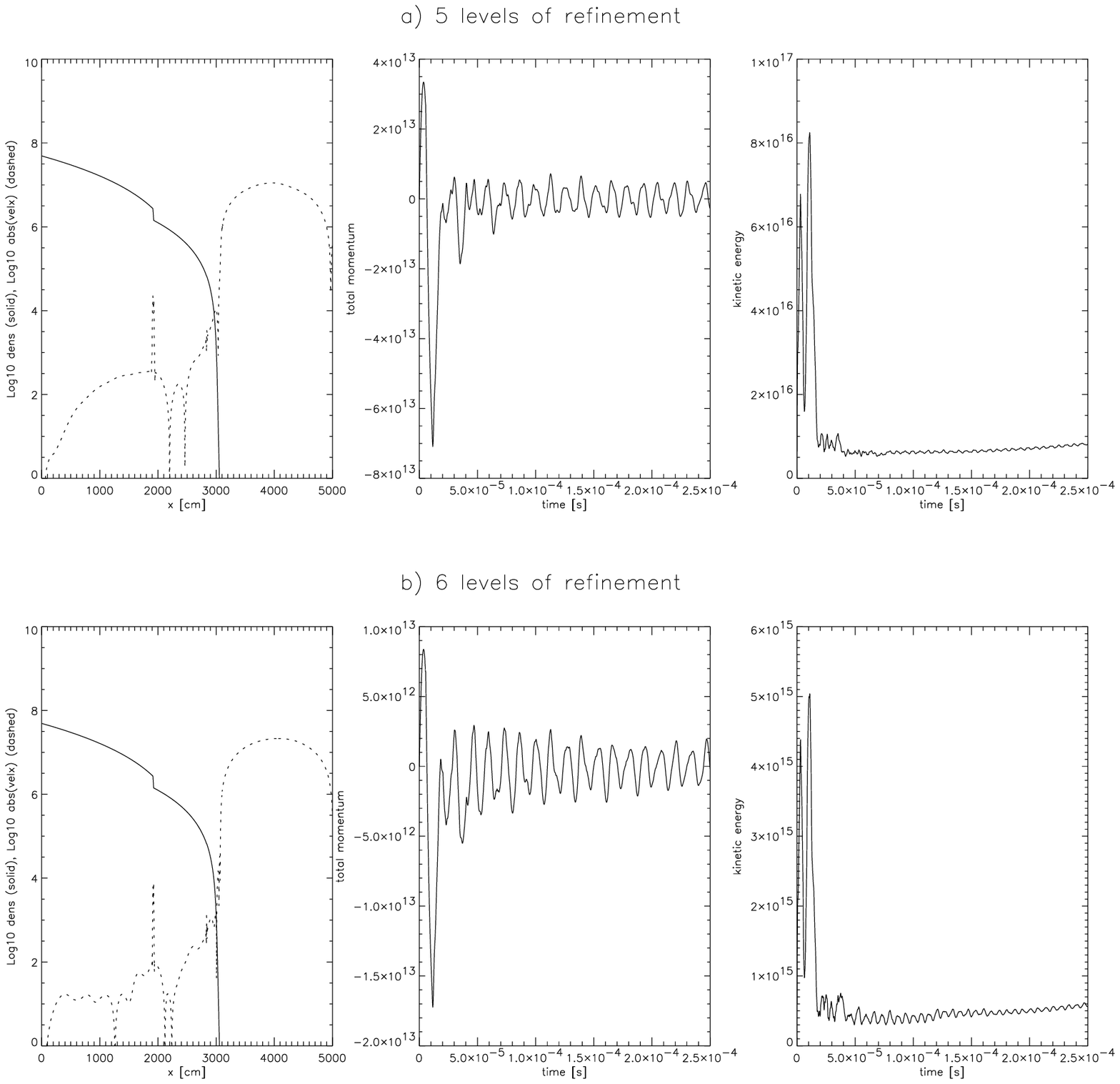}
\caption{\label{fig:xrb_res_study} Effect of the spatial resolution on
the velocity for the Kepler initial model.  Four different resolutions
are shown, a) 4.8 cm, b) 2.4 cm, c) 1.2 cm, and d) 0.6 cm.}

\end{figure}

\clearpage

\begin{figure}
\figurenum{\ref{fig:xrb_res_study}}
\epsscale{0.95}
\plotone{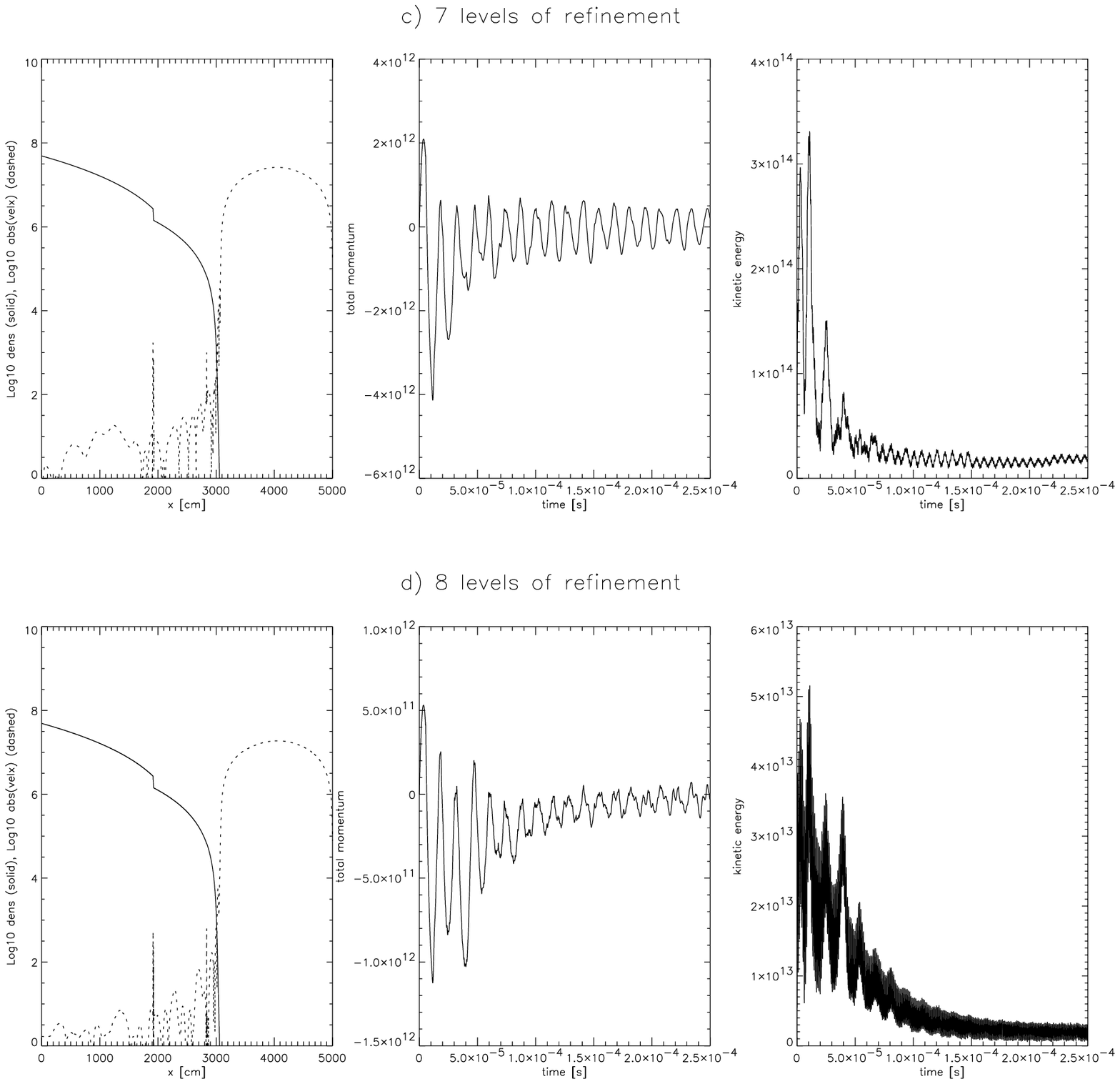}
\caption{cont.}

\end{figure}

\clearpage

\end{document}